\documentclass[12pt,letterpaper]{article}
\usepackage{epsfig,rotating,setspace,latexsym,amsmath,epsf,amssymb,amsfonts,bm,theorem,cite,caption,subcaption,enumerate,longtable,accents,algorithm,graphicx,epsf,authblk,epstopdf,url,color,multirow,algpseudocode,mathtools,comment,tabularx,comment,physics,graphicx,caption,subcaption}
\usepackage[short,c2]{optidef}
\usepackage[toc,title]{appendix}

\DeclarePairedDelimiter{\ceil}{\lceil}{\rceil}

\newtheorem{remark}{Remark}
\newtheorem{lemma}{Lemma}

\newtheorem{property}{Property}
\newenvironment{Proof}[1]{\medskip\par\noindent{\bf Proof:\,}\,#1}{{\mbox{\,$\blacksquare$}\par}}

\newcolumntype{Y}{>{\centering\arraybackslash}X}
\newcommand{\e}{{\mathbb{E}}}
\allowdisplaybreaks

\title{Cyclic Scheduler Design for Minimizing Age of Information in Massive Scale Networks \\ Susceptible to Packet Errors\thanks{This work is presented in part at IEEE Infocom, May 2024. This work is done when N.~Akar is on sabbatical leave as a visiting professor at University of Maryland, MD, USA, which is supported in part by the Scientific and Technological Research Council of T\"{u}rkiye  (T\"{u}bitak) 2219-International Postdoctoral Research Fellowship Program.}}

\author[1]{Sahan Liyanaarachchi}
\author[1]{Sennur Ulukus}
\author[2]{Nail Akar}

\affil[1]{\normalsize University of Maryland, College Park, MD, USA}
\affil[2]{\normalsize Bilkent University, Ankara, T\"{u}rkiye}

\begin{document}
\date{}
\maketitle

\vspace*{-1.0cm}

\begin{abstract}
   In multi-source status update systems, sources need to be scheduled appropriately to maintain timely communication between each of the sources and the monitor. A \emph{cyclic schedule} is an age-agnostic schedule in which the sources are served according to a fixed finite transmission pattern, which upon completion, repeats itself. Such a scheme has a low $O(1)$ runtime complexity, which is desirable in large networks. This paper's focus is on designing transmission patterns so as to be used in massive scale networking scenarios involving a very large number of sources, e.g., up to thousands of IoT sources, with service time requirements and weights being heterogeneous in nature. The goal is to minimize the weighted sum age of information (AoI), called weighted AoI, when transmitting users' packets over a channel susceptible to heterogeneous packet errors. The main tool we use is a stochastic modeling framework using either Markov chains (MC) or moment generating functions (MGF), by which we obtain the weighted AoI for a given transmission pattern, which is not straightforward in the presence of packet drops. Using this framework, we provide a lower bound on the weighted AoI for the particular case of two sources, and also an algorithm to attain this lower bound. Then, by using the same framework, we design a cyclic scheduler for general number of sources with reasonable complexity using convex optimization and well-established packet spreading algorithms, and comparatively evaluate the proposed algorithm and existing age-agnostic scheduling schemes for general number of sources (resp.~two sources) when the lower bound is not available (resp.~when it is available). We present extensive numerical results to validate the effectiveness of the proposed approach.
\end{abstract}

\section{Introduction} \label{sec:intro}
Need for timely status updates arises naturally in many remote monitoring and networked control systems \cite{gu2019timely, gupta2009networked, ayan2019age}. A multi-source status update system consists of a monitor which receives status updates from multiple sources through a shared communications channel \cite{Yates__SBR}. A common metric for quantifying timeliness in status update systems is the time average of the age of information (AoI) process, which we term as the mean AoI or average AoI \cite{kaul_etal_infocom12}, maintained at a certain device in the network, such as an intermediate node, a server, or the eventual destination, i.e., the monitor. In particular, the AoI process for source-$n$ denoted by $\Delta_n(t)$, $t\geq 0$, is defined as $\Delta_n(t)=t-g_n(t)$, where $g_n(t)$ is the time of generation of the freshest update of source-$n$ available at time $t$ to the particular device of interest \cite{RoyYates__AgeOfInfo_Survey, Kosta_AoI_2017, Sun_AoI_2019}. In this multi-source setting, status updates of some sources may be more critical than others, and therefore, weighted average of mean AoI values of the sources, termed weighted AoI, is often used as a system-wide metric for quantifying the timeliness of communication. To minimize the weighted AoI, source transmissions need to be scheduled appropriately. On the other hand, massive internet of things (IoT) networks are becoming commonplace that consist of a very large number of low cost and low power end users that report low frequency status updates to remote locations via 6G networks \cite{guo2021enabling} or low power wide area networks (LPWAN) such as LoRaWAN \cite{jouhari_etal_ST}. Development of scalable schedulers to be deployed in massive scale networks for weighted AoI minimization is the main scope of this paper. 

\begin{figure}[t]
    \centering
    \includegraphics[scale=0.75]{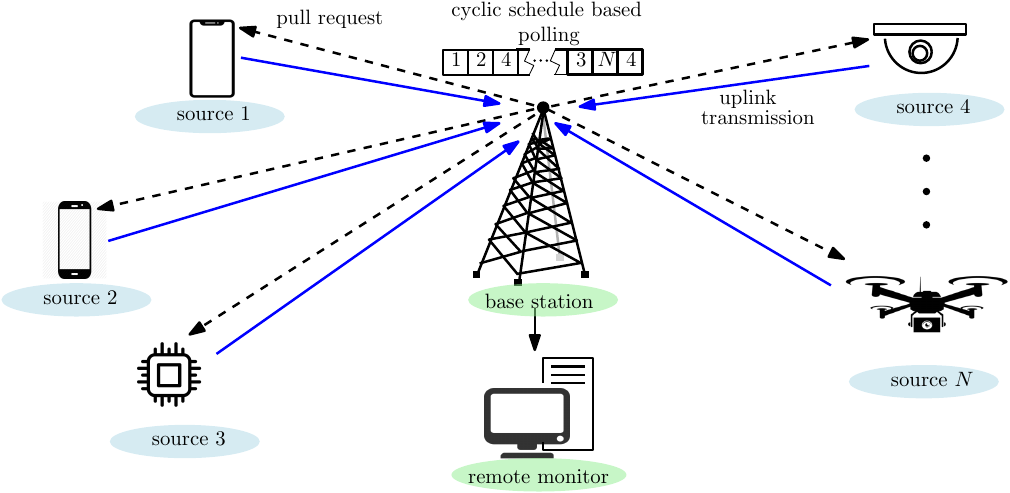}
    \caption{Pull-based status update system (Scenario A) involving a base station (BS) sending polling messages to $N$ users which in turn transmit their status update packets.}
    \label{fig:pull-based}
\end{figure}

The first representative scenario (called Scenario A) for a massive IoT network we consider in this paper is an error-prone cellular wireless network with a server or a base station (BS) collecting time-sensitive information from a large number of users that are spread out at random locations in the wireless network. Each user is associated with an information source process, whose sample values are to be transmitted to the server, which is in charge of forwarding these messages to a remote monitor, which may or may not be co-located with the server. Consequently, the terms ``user" and ``source" will be used interchangeably throughout the paper. In case of co-location, which is the scope of the current paper, AoI processes at both the server and the monitor are identical. However, these AoI processes may very well be different in the case of remotely located monitors, a situation which is left outside the scope of the current paper. Similar uplink scenarios are investigated in \cite{kadota2021wifresh, liu_etal_TCOM23, yu2024optimizing}, where the status update messages are generated according to a random process, or periodically. In this paper, we consider the \emph{generate-at-will} (GAW) model introduced in \cite{lts2015} which is a pull-based model where the server is to decide which source to collect the time-sensitive information from. Once the decision is made by the server, polling multiple access will be used by which the server sends a polling message (query) to the scheduled user which subsequently samples the corresponding scheduled source process, generates a status update packet, and transmits to the server in the uplink direction, while avoiding collisions. Although random access is generally used for channel access technology for timely status updates \cite{chen2020age, kadota2021age,chen2022age, yu2024optimizing}, we propose to use polling multiple access as in \cite{kadota2021wifresh} in order to prevent collisions that would stem from the cardinality of the number of sources attempting to transmit at the same time. Moreover, the polling message can be sent in the form of a wake-up radio signal upon reception of which the scheduled user wakes up and carries out the sampling and transmission operations before going back to sleep \cite{ghose2018mac, trotta2019bee, shiraishi2022query, shiraishi2024coexistence}. This prevents the unscheduled sources from wasting unnecessary energy, making it suitable for low power users in massive IoT networks. Fig.~\ref{fig:pull-based} illustrates Scenario A with $N$ sources. The users may use different modulation and coding schemes depending on their distance from the BS, and therefore, have heterogeneous packet service time (transmission time) requirements characterized with their first two moments. Moreover, packet drops are assumed to occur with different packet error rates for each user-to-server transmissions, and these statistics are a-priori known by the server. 

The second application scenario (Scenario B) also worth of investigation is the wireless downlink scenario where a base station sends time-sensitive information to a number of users through unreliable channels in a wireless broadcast network \cite{age_aware1} and we are interested in minimizing the weighted AoI maintained at the users. Wake-up radio can also accompany the system model of \cite{age_aware1} for energy efficiency purposes. In contrast to the first scenario, it is always the server which is sending time-sensitive information on the downlink and the feedback channel from the users back to the server may not exist. Therefore, maintaining a replica of the AoI processes at the server side may not even be possible, which presents another justification for the deployment of age-agnostic scheduling schemes. Both scenarios A and B can be addressed with the proposed methods of the current paper. 

Two main approaches stand out in the literature for scheduling in multi-source status update systems. In age-aware scheduling, there are schedulers that make use of the framework of Markov decision processes (MDP) and dynamic programming (DP) with proven optimality \cite{hsu_etal_isit17, liu_etal_TCOM23, vini_etal_TVT24}. However, computing the optimal policy using DP becomes prohibitive for large values of $N$ since the state-space of the MDP grows exponentially with $N$. Hence,  curse of dimensionality inherent to MDP models \cite{gast2010optimization} hinders one from developing schedulers for massive connectivity scenarios. The alternative age-aware approach is the sub-optimal index-based scheduling such as the ones proposed in \cite{age_aware1, age_aware2,age_aware3}, for which the scheduling policy is relatively easier to develop off-line. However, their runtime implementation still requires the server to continuously keep track of the AoI of the sources, and $O(N)$ computational complexity to choose the source with the highest index. Although far lower than the complexity of DP counterparts, the runtime computational burden of such algorithms is one of the limitations of age-aware schemes for massive IoT networks. Moreover, age-aware schemes fail to extend to scenarios in which the server cannot always perfectly maintain a replica of the actual AoI. This may happen since in the first scenario, monitors may be located away from the server introducing additional delays and losses in the forward and reverse channels from the server to monitors, and in the second scenario, ACK/NACK messages may not be sent from the users back to the server. In the alternative age-agnostic schemes, and in particular cyclic scheduling \cite{eywa, gau_ciss24}, the server does not need to keep track of the ages of the individual sources while making scheduling decisions. Instead, the server develops a cyclic schedule, i.e., a fixed pattern of user transmissions which repeats itself, with an off-line algorithm using a-priori known statistical information. Once the schedule is developed, its runtime complexity is $O(1)$ which makes it an exceptionally suitable scheme for massive IoT networks. 

The main scope of this paper is the off-line development of low complexity cyclic schedules for weighted AoI minimization that can scale to very large number of users encountered in massive IoT settings. In this work, we extend our previous work on cyclic scheduling \cite{gau_ciss24} in three directions: First, we provide an analytical method to compute the weighted AoI in the same setting as in \cite{gau_ciss24} but also taking into consideration the case of heterogeneous packet errors, whereas no packet errors were assumed in \cite{gau_ciss24}. Second, \cite{gau_ciss24} shows in the absence of packet errors and for the specific case of two sources that, the optimal cyclic scheduler is of the form $(1,\Theta)$, which represents a cyclic schedule where one scheduling instance of one of the sources is followed by $\Theta$ scheduling instances of the other source. In this paper, we show that for two heterogeneous sources and in the presence of packet errors, a near-optimal cyclic scheduler is in the form of a mixture of $(1,\Theta)$ and $(1,\Theta+1)$ cyclic schedulers, and we provide a novel algorithm for its construction. Although the $N=2$ case is definitely not a large-scale scenario, having the optimum solution in hand for this special case enables us to compare various heuristic approaches against the optimum solution. Third, \cite{gau_ciss24} proposes the IS (insertion search) heuristic algorithm for $N>2$ but its computational complexity exceeding $O(N^5)$ makes it unfit for massive connectivity scenarios. In the current paper, we propose a heuristic algorithm which is entirely different than IS and which can scale to massive IoT scenarios. A preliminary version of the current paper can be found in \cite{akar_etal_infocom24}.  

Our main contributions are summarized as follows:
\begin{itemize}
    \item We provide two novel analytical frameworks, MC-based and MGF-based, for the weighted AoI computation of a given cyclic schedule, both of which are substantially different than the method described in \cite{gau_ciss24}, since the behavior of the constructed cyclic schedule will no longer be cyclic due to packet drops. MC-based and MGF-based approaches both have their own merits which will become apparent in Section~\ref{sec:2source}.
    \item For the special case of $N=2$, we prove that given any $\epsilon>0$, our algorithm constructs a cyclic scheduler whose weighted AoI is within $\epsilon$ of the actual optimum. Therefore, we establish a lower bound for the weighted AoI for the $N=2$ case, which can be used as a baseline for comparative evaluation of age-agnostic algorithms in this special case.
    \item We propose algorithms for the construction of a cyclic schedule appropriate for massive IoT scenarios with the goal of weighted AoI minimization, which are based on convex optimization and well-established packet spreading algorithms. We show through numerical results and simulations that our algorithms outperform existing age-agnostic multi-source schedulers in systems involving very large numbers of users, in terms of weighted AoI minimization. Moreover, their relatively low off-line complexity as well as $O(1)$ runtime complexity makes them a good fit for massive connectivity scenarios.
\end{itemize}

The remainder of the paper is organized as follows. Section~\ref{sec:related} overviews the related work. Section~\ref{sec:systemmodel} presents the system model. Section~\ref{sec:analysis} develops the analytical method to obtain the weighted AoI when cyclic scheduling is employed. Section~\ref{sec:2source} derives the optimum cyclic scheduler in closed form for the case of two sources. Section~\ref{sec:multi} presents the proposed scalable scheduling algorithm for general number of sources. Section~\ref{sec:other} describes the existing baseline age-agnostic schemes along with the modifications we have introduced in this work. Section~\ref{sec:numerical} presents extensive numerical results. Section~\ref{sec:conclusions} concludes the paper. 

\section{Related Work}\label{sec:related}
AoI metric was first proposed in a single-source, single-server queuing scenario \cite{kaul_etal_infocom12}. A survey of existing research on AoI and its applications can be found in \cite{RoyYates__AgeOfInfo_Survey, Kosta_AoI_2017, Sun_AoI_2019}. Several AoI stochastic analysis methods have been developed in the literature.  For random arrivals (RA), where status updates take place according to a random process, e.g., such as the Poisson process, the stochastic hybrid systems (SHS) approach of \cite{Yates__SBR} is generally used in the literature for obtaining the mean AoI, whereas SHS can also be used to obtain the moment generating function (MGF) and also the higher moments of AoI in certain scenarios \cite{yates_mgf, moltafet_etal_tcom22}. On the other hand, for multi-source GAW systems (and also for a special multi-source RA system with one single packet buffer), an alternative absorbing Markov chain (AMC) based method is proposed to find the distribution of AoI in \cite{akar_gamgam_comlet23}, which requires a-priori information about the service time distributions. The current paper is different than \cite{akar_gamgam_comlet23}, since we only focus on the average AoI in this paper (as opposed to its distribution), which only depends on the first two moments of the per-source service times (and not on their distributions), and also on the packet error probabilities. 

Age-aware scheduling for AoI minimization has extensively been studied in the literature. \emph{Maximum age first (MAF)} policy \cite{maf1, maf2, maf3}, where the source with the highest instantaneous age is scheduled, and \emph{maximum-age-difference drop (MAD)} policy \cite{mad}, where the source which would result in the maximum drop in age is scheduled, have been proposed in the literature for homogeneous multi-source settings. For heterogeneous scenarios, max-weight and Whittle-index policies \cite{age_aware1, age_aware2, age_aware3} have been proposed that are shown to perform very well despite the lack of proof of optimality in general scenarios. Timeliness of communication in federated learning (FL), which was addressed in \cite{buyukates_2021}, is another domain where age-aware scheduling policies have been utilized. Further,  \cite{yang_etal_ICASSP20} proposes an age-aware scheduling policy for FL which also takes into account the instantaneous channel qualities.

Age-aware scheduling policies require the scheduler to continuously track the age of the sources, and therefore, can introduce a significant communication overhead, especially in channels susceptible to packet drops. Moreover, in open-loop communication systems where feedback on packet drops is absent, age-aware scheduling may not even be feasible. Therefore, age-agnostic cyclic scheduling has recently appeared as a low cost open-loop solution compared to age-aware scheduling \cite{cyclic1, cyclic2, eywa, gau_ciss24}. A framework called \emph{Eywa} was recently introduced in \cite{eywa}, where the goal is to construct \emph{almost uniform cyclic schedulers (AUS)}, which is a special class of cyclic schedulers designed to distribute the scheduling instances of a given source as uniformly as possible within the cycle. \emph{Eywa} works in a discrete time setting and assumes that all sources have the same deterministic service times with heterogeneous packet errors. However, in real systems, service time requirements may be different across the sources, for which case  reference \cite{gau_ciss24} obtains the optimal cyclic scheduler that minimizes the weighted AoI for two heterogeneous sources, and also develops a heuristic scheduler for the case of several sources, but in the absence of packet errors. The focus of our current paper is on age-agnostic scheduling in a massive connectivity setting, and also in the presence of packet errors.

\section{System Model} \label{sec:systemmodel}
Consider the status update system shown in Fig.~\ref{fig:pull-based}, consisting of $N$ sources indexed by $n = 1, 2, \dots, N$. The sources are scheduled according to a cyclic schedule, where each source, once scheduled, samples its associated random process and generates a data packet which is transmitted through the shared channel. We consider a random delay channel which is prone to packet drops, where the $n$th source is subject to a packet drop probability $p_n$ and the packet success probability $u_n=1-p_n$. The random delay (service time) experienced by the $n$th source is denoted by the i.i.d.~random variable $S_n$ with mean $s_n =\e[S_n]$, second moment $q_n=\e[S_n^2]$, variance $v_n = q_n-s_n^2$, and squared coefficient of variation (scov) $c_n=\frac{v_n}{s_n^2}$. The fixed wake-up signal transmission time for querying purposes is assumed to be much shorter compared to the information packet service times in both Scenarios A and B. Once a scheduled source has finished its transmission, a new sample from the next source in schedule will be immediately sampled and transmitted. The transmitter (a user in the case of Scenario A and the server in the case of Scenario B) has no information whether a packet was successfully received or was dropped due to the open-loop nature of the system.

\begin{figure}[t]
    \centering
    \includegraphics[width=0.65\textwidth]{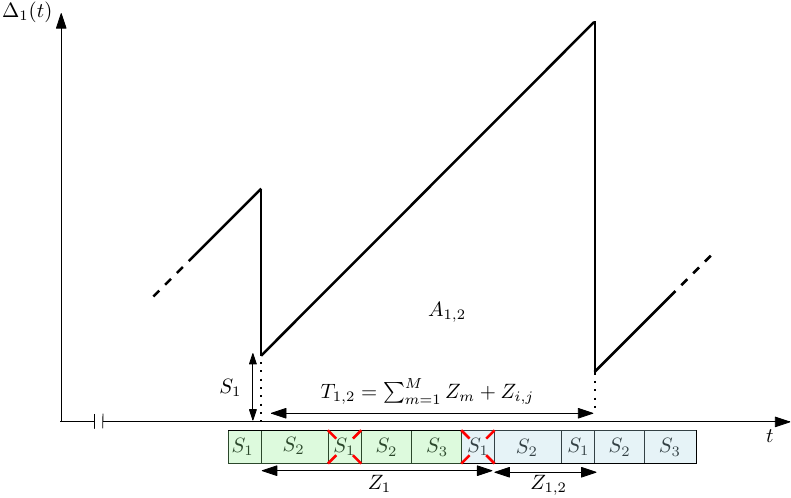}
    \caption{AoI graph of source-$1$ for the pattern $P=[1,2,1,2,3]$ showing a realization of the area $A_{1,2}$ when $M = 1$.}
    \label{fig:aoi}
\end{figure}

The cyclic schedule is characterized by a pattern $P = [P_0, P_1, \ldots, P_{K-1}]$, which is a row vector of size $K$ such that $P_k \in \{1,2,\ldots,N \}$. Based on the pattern $P$, the cyclic scheduler selects source-$P_k$ for transmission at the scheduling instant $k+iK$, $i \in \mathbb{Z^+}$. As an example, let $N=3$ and $P = [1,2,1,2,3 ]$. In this case, the cyclic scheduler will transmit the following sequence of source packets: $1,2,1,2,3,1,2,1,2,3,1,2,1,2,3,\ldots$. A pattern is said to be feasible if each of the $N$ sources appears at least once in the pattern (otherwise, AoI of one or more sources will be unbounded). Thus, $K\geq N$. Let $\Delta_n$ denote the steady-state random variable for the associated age process, denoted by $\Delta_n(t)$, observed at the server for source-$n$, $1 \leq n \leq N$. The age process of source-$n$ increases linearly with a unit slope with respect to time and upon successful reception (without packet drop) of a data sample from source-$n$, the age drops to the value of the delay experienced by that particular data packet that was successfully received. Fig.~\ref{fig:aoi} depicts the evolution of the age process of source-$1$ for the above pattern $P$, where failed source-$1$ transmissions are crossed out in red. The per-source service times are heterogeneous but deterministic in Fig.~\ref{fig:aoi}. As shown in Fig.~\ref{fig:aoi}, $\Delta_1(t)$ continues to increase linearly during the transmission of samples from other sources as well as during failed source-$1$ transmissions, and drops upon successful reception of a source-$1$ sample. We define the weighted AoI as the mean of the random variable $\Delta = \sum_{n=1}^N w_n \Delta_n$, i.e.,
\begin{align}
    \mathbb{E} [\Delta]  =  \sum_{n=1}^N w_n \mathbb{E} [\Delta_n], 
\end{align}
where the normalized source weights $w_n$, $\sum_{n=1}^N w_n=1$, are used to prioritize the sources. 

The main goal of this work is to design a well-performing cyclic scheduler, i.e., a transmission pattern $P$, with an attempt to minimize the weighted AoI by taking packet drop probabilities into consideration. As the first step towards realizing this goal, we present two separate approaches for analytically obtaining the weighted AoI of a cyclic scheduler. 

\section{AoI Analysis} \label{sec:analysis}
Even though we start off with a cyclic schedule, due to packet drops, the actual schedule of successful transmissions will no longer behave according to the constructed schedule. Thus, weighted AoI computation is a difficult task even for two sources. In this section, we present two methods, namely MC- and MGF-based methods, which can be used to compute the average AoI of a cyclic scheduling scheme. Each method has its merits, where the former one will be used to construct a near-optimal scheduler for two heterogeneous sources and the latter one will be employed in the design of large-scale cyclic schedulers.

\subsection{Markov Chain (MC) Formulation}
\begin{figure}
    \centering
    \includegraphics[width=0.9\textwidth]{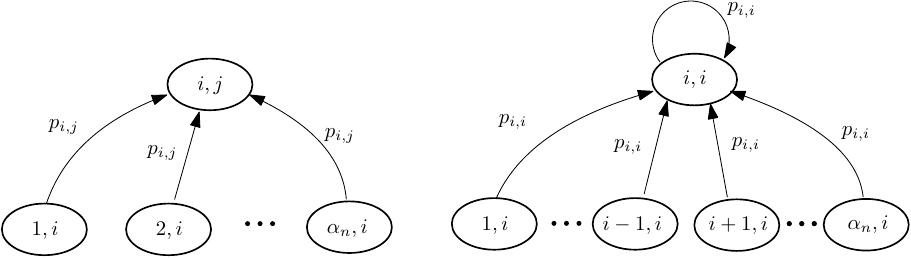}
    \caption{Partial state transition diagram for $\alpha_n>2$.}
    \label{fig:u-MC}
\end{figure}

To derive the AoI expression in here, we first consider a single source (say source-$n$) and characterize its AoI process using a Markov chain (MC) formulation. Let $\alpha_n$, $n=1,\ldots,N$ denote the number of times that source-$n$ appears in the pattern $P$. Note that $\alpha_n > 0$, for all $n$, since otherwise it will not be a feasible schedule (weighted AoI will be unbounded). There are $\alpha_n$ locations within the pattern $P$ where the AoI of source-$n$ can drop. Therefore, based on where two consecutive AoI drops occur relative to the pattern $P$, we can define $\alpha_n^2$ states for the MC. Let $(i,j)$ for $i,j \in \{1,2, \ldots, \alpha_n\}$ denote an AoI cycle starting from the $i$th scheduling instance of source-$n$ and ending in the $j$th scheduling instance of source-$n$. Note that $(i,j)$ represents the $i$th and the $j$th scheduling instances of source-$n$ relative to the pattern $P$. Each state transition occurs after successfully (without packet drop) transmitting a  sample from source-$n$. Fig.~\ref{fig:u-MC} shows a partial state transition diagram for the MC.

As shown in Fig.~\ref{fig:u-MC}, state $(i,j)$ is only directly accessible by states of the form $(k,i)$ and they all share the same transition probability $p_{i,j}$. Only states of the form $(i,i)$ have self transitions. The transition probabilities are given as,
\begin{align}
    p_{i,j} &= \frac{(1-p_n)p_n^{j-i-1}}{1-p_n^{\alpha_n}}, & \text{if } j>i, \\
    p_{i,j} &= \frac{(1-p_n)p_n^{j+\alpha_n-i-1}}{1-p_n^{\alpha_n}}, & \text{if }  i\geq j.
\end{align}
Since these are finite-state MCs with no absorbing states for $p_n>0$, they are positive recurrent and irreducible. Since they contain self-loops, they are aperiodic and hence are ergodic. Let $\pi=\{\pi_{(i,j)}\}$  for $i,j \in \{1,2, \dots ,\alpha_n\}$ denote the stationary distribution of the MC. Then, the stationary distribution takes the following form,
\begin{align}
    \pi_{i,j} = \frac{p_{i,j}}{\alpha_n}.
\end{align}
To compute the average AoI, we partition the AoI (i.e., age) graph based on the states of the MC. Let $A_{i,j}$ denote the area under the age curve  when the two consecutive age drops are placed at the $i$th and $j$th scheduling instances of the cycle (see, for an example, $A_{1,2}$ in Fig.~\ref{fig:aoi}). Let $T_{i,j}$ denote the  time spent for the aforementioned age drops. Since the MC is ergodic, the average AoI of source-$n$ is given by,
\begin{align}
    \mathbb{E}[\Delta_n] = \frac{\sum_{i=1}^{\alpha_n}\sum_{j=1}^{\alpha_n}\pi_{i,j}\e[A_{i,j}]}{\sum_{i=1}^{\alpha_n}\sum_{j=1}^{\alpha_n}\pi_{i,j}\e[T_{i,j}]}. \label{eqn:aoi_stat}
\end{align}

Therefore, to compute the average AoI, we need to find $\e[A_{i,j}]$ and $\e[T_{i,j}]$. $A_{i,j}$ consists of the segments of the age curve with consecutive age drops happening at the $i$th and $j$th scheduling instances of source-$n$. Suppose an AoI drop happens at the $i$th scheduling instance, then the next AoI drop may happen at the $j$th scheduling instance after going through multiple rounds of the entire pattern $P$. This is illustrated in Fig.~\ref{fig:aoi} considering the average AoI of source-$1$. Let $M$ denote the number of rounds of the entire pattern $P$ that has elapsed before the next successful transmission of a source-$n$ sample occurring at the $j$th scheduling instance. For example, if the next successful transmission occurs in the second round, then $M=1$. Then, $M=\hat{M}-1$, where $\hat M$ is a geometric random variable with  parameter $p^{\alpha_n}$ (i.e., $\hat{M} \sim \text{Geom}(p^{\alpha_n})$). Let $Z_{i,j}$ denote the time duration elapsed starting from the beginning of the $(M+1)$th round to the AoI drop occurring at the $j$th scheduling instance of source-$n$. Let $Z_m$ denote the time duration of the $m$th round. Then, $\e[A_{i,j}]$ can be written as,
\begin{align}
    \mathbb{E}[A_{i,j}] &= \frac{\e\left[\left(2S_n+\sum\limits_{m=1}^MZ_m+Z_{i,j}\right)\left(\sum\limits_{m=1}^MZ_m+Z_{i,j}\right)\right]}{2} \nonumber\\
    &=\frac{\tilde{m}\hat{s}^2+\overline{m}(2s_n\hat{s}+\hat{v})+\tilde{z}_{i,j}+2\overline{z}_{i,j}(\overline{m}\hat{s}+s_n)}{2},\label{eqn:ai}
\end{align}
where $\hat{s}=\e[Z_m]=\sum_{k=1}^N\alpha_ks_k$, $\hat{v} =\sum_{k=1}^N\alpha_kv_k$,  $\e[M] = \overline{m}$, $\e[M^2] = \tilde{m}$, $\e[Z_{i,j}] = \overline{z}_{i,j}$ and $\e[Z_{i,j}^2] = \tilde{z}_{i,j}$. Similarly, $\e[T_{i,j}]$ can be written as,
\begin{align}
    \e[T_{i,j}] =\e\left[\sum_{m=1}^MZ_m+Z_{i,j}\right] = \overline{m}\hat{s}+\overline{z}_{i,j}. \label{eqn:T}
\end{align}
Then, by substituting (\ref{eqn:ai}) and (\ref{eqn:T}) into (\ref{eqn:aoi_stat}), we find the average AoI of source-$n$ as follows,
\begin{align}
    \e[\Delta_n]= \frac{\tilde{m}\hat{s}^2+\overline{m}(2s_n\hat{s}+\hat{v})+\sum_{i,j}\pi_{i,j}\tilde{z}_{i,j}}{2(\overline{m}\hat{s}+\sum_{i,j}\pi_{i,j}\overline{z}_{i,j})}+\frac{(\overline{m}\hat{s}+s_n)\sum_{i,j}\pi_{i,j}\overline{z}_{i,j}}{\overline{m}\hat{s}+\sum_{i,j}\pi_{i,j}\overline{z}_{i,j}}. \label{eqn:u1_aoi}
\end{align}

\subsection{Moment Generating Function (MGF) Formulation}\label{sec:mgf_analysis}
In this section, we present an alternative way to derive the weighted AoI of a cyclic schedule using moment generating functions (MGF). The reason for the development of this alternative method is due to the fact that the age expression obtained by this method can be easily fashioned into an optimization problem that can be used to construct well-performing heuristic solutions for the development of cyclic schedules for a large number of sources.

Let $S_{n,k}$ denote the service time of the $k$th successful transmission of source-$n$ and let $\tilde{S}_{n,k}$ denote the time duration between the end of the $k$th successful transmission and the beginning of the $(k+1)$th successful transmission of source-$n$. Note that $\tilde{S}_{n,k}$ is the sum of the service times of all packets generated from sources other than source-$n$ between two consecutive successful transmissions of source-$n$, denoted by $\tilde{S}^o_{n,k}$, and also of the service times of unsuccessful packets belonging to source-$n$ itself, denoted by $\tilde{S}^u_{n,k}$. Clearly, $\tilde{S}_{n,k} = \tilde{S}^o_{n,k} + \tilde{S}^u_{n,k}$. Let the random variable $\tilde{S}_n$ (steady-state random variable associated with the random process $\tilde{S}_{n,k}$ as $k \rightarrow \infty$) have mean $\tilde{s}_n =\mathbb{E} [\tilde{S}_n]$, second moment $\tilde{q}_n = \mathbb{E} [\tilde{S}_n^2]$, variance $\tilde{v}_n = \tilde{q}_n - \tilde{s}_n^2$ and scov $\tilde{c}_n = \frac{\tilde{v}_n}{\tilde{s}_n^2}$. Based on \cite{gau_ciss24}, the mean AoI for source-$n$ can be written in terms of the first two moments of the random variables $S_n$ and $\tilde{S}_n$ as,
\begin{align}
    \mathbb{E}[\Delta_n]  & = \frac{2 s_n^2 + 4 s_n \tilde{s}_n + q_n + \tilde{q}_n}{2(s_n + \tilde{s}_n)} \label{exp_2mom} \\
    & = \frac{s_n^2 (c_n + 3) +\tilde{s}_n^2(\tilde{c}_n + 1) + 4 s_n \tilde{s}_n }{2(s_n + \tilde{s}_n)}. \label{exp_scov}
\end{align}

To find the average AoI of source-$n$, we need to compute the parameters $\tilde{s}_n$ and $\tilde{c}_n$. Let $P_{n,k}$, $k=0,1,\ldots,\alpha_n-1,$ denote the sub-pattern between the $k$th and $(k+1)$th appearances of source-$n$ within $P$, excluding the end points. Let $\alpha_{n,k,m}$ denote the number of times source-$m$ appears in the sub-pattern $P_{n,k}$. Let us denote by $H_{n,k}$, the random variable corresponding to the sum of the service times of all sources contained within the sub-pattern $P_{n,k}$. Then, the MGF of $H_{n,k}$, denoted by $G_{n,k}(s) $, is obtained as follows,
\begin{align}
    G_{n,k}(s) & = \mathbb{E} [e^{sH_{n,k}}] = \prod\limits_{m \neq n} G_m(s)^ {\alpha_{n,k,m}}.
\end{align}
Let $\tilde{H}_{n,k}$ denote the time between two consecutive successful transmissions of source-$n$ (excluding the end points) given that the first successful transmission occurs at the $k$th appearance of source-$n$ in the pattern $P$ and let
$\tilde{G}_{n,k}(s) = \mathbb{E} [e^{s \tilde{H}_{n,k}}]$ denote its MGF. From the total law of expectation, we have,
\begin{align}
    \frac{\tilde{G}_{n,k}(s)}{u_n} =  G_{n,k}(s) 
    &+  p_n G_n(s)G_{n,k}(s)  G_{n,k+1}(s)  +  p_n^2 G_n(s)^2 G_{n,k}(s)  G_{n,k+1}(s) G_{n,k+2}(s)  +  \ldots \nonumber\\
    &  +  p_n^{\alpha_n-1} G_n(s)^{\alpha_n-1} {\prod_{m=1}^{\alpha_n} G_{n,m-1}(s)}   +  p_n^{\alpha_n} G_n(s)^{\alpha_n}  G_{n,k}(s) {\prod_{m=1}^{\alpha_n} G_{n,m-1}(s)} \nonumber\\
    &  +  p_n^{\alpha_n} G_n(s)^{\alpha_n +1} G_{n,k}(s) G_{n,k+1}(s) {\prod_{m=1}^{\alpha_n} G_{n,m-1}(s)}  +  \ldots \label{mgf75}
\end{align}
where $G_{n,k}(s)=G_{n,k+\alpha_n}(s)$, $0 \leq k \leq \alpha_n$, by convention, and $u_n=1-p_n$. A successful source-$n$ packet belongs to a transmission opportunity at appearance-$k$, $k=0,1,\ldots,\alpha_n$, uniformly. Therefore, the MGF $\tilde{G}_{n}(s)$ of the random variable $\tilde{S}_n$ is written as,
\begin{align}
    \tilde{G}_{n}(s) &= \frac{1}{\alpha_n} \sum_{k=1}^{\alpha_n} \tilde{G}_{n,k-1}(s). \label{nail76}     
\end{align}
Let $s_{n,k}$ and $q_{n,k}$ represent the mean and the second moment of $H_{n,k}$. By differentiating \eqref{mgf75} with respect to the variable $s$ and evaluating the result at $s=0$, we obtain,
\begin{align}
    \tilde{s}_n &=  \mathbb{E} [\tilde{S}_n] = \tilde{G}_n'(0) = \frac{1}{\alpha_n} \sum_{k=1}^{\alpha_n} \tilde{G}'_{n,k-1}(0) \\
    & = \frac{1}{\alpha_n} \sum_{k=1}^{\alpha_n} \left( \frac{p_n s_n}{u_n} + \frac{\sum_{j=1}^{\alpha_n} p_n^{j-1} s_{n,k+j-1}}{1-p_n^{\alpha_n}} \right)\\
    & =\frac{1}{u_n} \left( p_n s_n +  \frac{1}{\alpha_n} \sum_{k=1}^{\alpha_n} s_{n,k-1} \right).
    \label{nail77}           
\end{align}
Again, in the equations above, we used $s_{n,k}=s_{n,k+\alpha_n}$, $0 \leq k \leq \alpha_n-1$, by convention. To obtain $\tilde{q}_n$, we note that \eqref{mgf75} can be simplified as follows,
\begin{align}
    \tilde{G}_{n,k}(s) & =  
    \frac{u_n \sum_{j=1}^{\alpha_n} p_n^{j-1} G_n(s)^{j-1} \prod_{l=1}^j G_{n,k+l-1}(s) }{1 - p_n^{\alpha_n}G_n(s)^{\alpha_n} {\prod_{m=1}^{\alpha_n} G_{n,m-1}(s)}}.                  
\end{align}
Ignoring the higher order terms (since they will not have any effect on $\tilde{G}_{n,k}''(0)$), we write,
\begin{align}
    G_n(s) & = 1 + s_n s + \frac{q_n}{2} s^2 + O(s^3), \\
    G_{n,k}(s) & = 1 + s_{n,k} s + \frac{q_{n,k}}{2} s^2 + O(s^3),
\end{align}
from which we can write, 
\begin{align}
    \tilde{G}_{n,k}(s) & = \frac{(1-p_n^{\alpha_n}) + a_{n,k} s + {b_{n,k}} s^2 + O(s^3)}{(1-p_n^{\alpha_n}) + c_{n,k} s + d_{n,k} s^2 + O(s^3)}, \label{eqn:gnk_num}
\end{align}
where the coefficients $a_{n,k}$, $b_{n,k}$, $c_{n,k}$ and $d_{n,k}$ can be found numerically, e.g., by applying the convolution operator for the product of MGFs. Then, by differentiating \eqref{eqn:gnk_num} twice and evaluating at $s=0$, the $O(s^3)$ terms will vanish, yielding $\tilde{q}_n$ as follows,
\begin{align}
    \tilde{q}_n &=  \mathbb{E} [\tilde{S}_n^2] = \tilde{G}_n''(0) = \frac{1}{\alpha_n} \sum_{k=1}^{\alpha_n} \tilde{G}''_{n,k-1}(0) \\
    & = \frac{1}{\alpha_n} \sum_{k=1}^{\alpha_n} \left( \frac{2c_{n,k}(c_{n,k}-a_{n,k})} {(1-p_n^{\alpha_n})^2} +
    \frac{2(b_{n,k}-d_{n,k})} {(1-p_n^{\alpha_n})} \right). \! 
    \label{nail79}           
\end{align}
Substituting the values of $\tilde{s}_n$ and $\tilde{q}_n$ obtained in \eqref{nail77} and \eqref{nail79}, respectively, into \eqref{exp_2mom}, gives the average AoI of source-$n$.

\section{Cyclic Scheduling for Two Sources ($N=2$)}\label{sec:2source}
In this section, we analyze the average weighted age expression obtained by the MC formulation for two heterogeneous sources, and present algorithms to construct two source cyclic schedulers which we subsequently prove to be near optimal.

\subsection{Near Optimal Two Source Scheduler (NOTS)}
Let $\bm{r}=\{r_1,r_2,\dots,r_{\alpha_1}\}$ represent the placement vector of the schedule with respect to source-$1$, where $r_i$ is the number of source-$2$ scheduling instances between the $i$th and the $(i+1)$th scheduling instances of source-$1$. Note that $\alpha_2 = \sum_ir_i$. Any scheduling pattern $P$ corresponding to a two source cyclic scheduler, can be uniquely represented by this placement vector $\bm{r}$. For example, the pattern $P=[1,2,2,2,1,2,1,2]$ would correspond to placement vector $\bm{r}=\{3,1,1\}$. Let $a=\frac{\alpha_2}{\alpha_1}$, $s=as_2+s_1$ and $v=av_2+v_1$. Considering the average AoI of source-$1$, based on the symmetric nature of the stationary distribution, we can further simplify \eqref{eqn:u1_aoi} using the following,
\begin{align}
     \sum_{i,j}\pi_{i,j}\overline{z}_{i,j} &= s\frac{(1-p_1)}{(1-p_1^{\alpha_1})}\sum_{i=1}^{\alpha_1}ip_1^{i-1}, \label{eqn:pi_z}
     \end{align} 
     and
     \begin{align}
    \sum_{i,j}\pi_{i,j}\tilde{z}_{i,j} &=\frac{(1-p_1)}{(1-p_1^{\alpha_1})}\sum_{i=1}^{\alpha_1}\left[iv+i^2s^2\right]p_1^{i-1}+s_2^2\frac{(1-p_1)}{\alpha_1(1-p_1^{\alpha_1})}\sum_{i=1}^{\alpha_1}(\tilde{r}(i)-\alpha_1a^2i^2)p_1^{i-1}. \label{eqn:pi_zz}
\end{align}
In  (\ref{eqn:pi_zz}), $\Tilde{r}(i)$ is defined as follows,
\begin{align}
    \Tilde{r}(i) = \sum_{j=1}^{\alpha_1}\left(\sum_{k=j}^{j+i-1}\gamma_k\right)^2,
\end{align}
where $\bm{\gamma}= \{r_1,r_2, \dots,r_{\alpha_1},r_1,r_2, \dots,r_{\alpha_1}\}$.
The $\tilde{r}(i)$ term is simply the sum of squared sum of $i$ consecutive elements in the placement vector. For example, $\tilde{r}(1)$, $\tilde{r}(2)$ and $\tilde{r}(3)$  are,
\begin{align}
    \tilde{r}(1) &= r_1^2 + r_2^2 + \dots +r_{\alpha_1}^2,\\
    \tilde{r}(2) &= (r_1+ r_2)^2 + (r_2+ r_3)^2 +\dots+(r_{\alpha_1}+r_1)^2,\\
    \tilde{r}(3) &= (r_1+r_2+r_3)^2 + (r_2+r_3+r_4)^2+\dots + (r_{\alpha_1}+r_1+r_2)^2.
\end{align}
Substituting (\ref{eqn:pi_z}) and (\ref{eqn:pi_zz}) into (\ref{eqn:u1_aoi}) yields average AoI of source-$1$ as,
\begin{align}
    \e[\Delta_1] &= \frac{(1+p_1)}{2(1-p_1)}s + \frac{v}{2s}+s_1 + s_2^2\frac{(1-p_1)^2}{2s\alpha_1(1-p_1^{\alpha_1})}\sum_{i=1}^{\alpha_1}\left(\tilde{r}(i)-\alpha_1a^2i^2\right)p_1^{i-1}. \label{eqn:aoi_1}
\end{align}
If $\bm{l}=\{l_1,l_2,\dots,l_{\alpha_2}\}$ represents the placement vector of the schedule with respect to source-$2$, then the average AoI of source-$2$, $\e[\Delta_2]$, can be obtained by replacing $a$, $s$, $v$, $s_1$, $s_2$, $p_1$, and $\alpha_1$ in \eqref{eqn:aoi_1} with $\frac{1}{a}$, $\frac{s}{a}$, $\frac{v}{a}$, $s_2$, $s_1$, $p_2$, and $\alpha_2$, respectively, to write,
\begin{align}
    \e[\Delta_2] &= \frac{(1+p_2)}{2a(1-p_2)}s + \frac{v}{2s}+s_2 + s_1^2\frac{a(1-p_2)^2}{2s\alpha_2(1-p_2^{\alpha_2})}\sum_{i=1}^{\alpha_2}\left(\tilde{l}(i)-\frac{\alpha_2i^2}{a^2}\right)p_2^{i-1}. \label{eqn:aoi_2}
\end{align}
\begin{remark}
    When $p_1=p_2=0$, equations \eqref{eqn:aoi_1} and \eqref{eqn:aoi_2} reduce to the average AoI expression obtained in \cite{gau_ciss24} which is a special case of the model studied here. 
\end{remark}

To construct the optimal cyclic scheduler, we need to find the optimal placement vector $\bm{r}$ that minimizes weighted AoI, $w_1\e[\Delta_1]+w_2\e[\Delta_2]$. To proceed with our analysis, for fixed $\alpha_1$ and $\alpha_2$, we first find the optimal placement vector that minimizes \eqref{eqn:aoi_1}. By relaxing the integer constraint on the placement vector, it follows that the minimum is achieved when $r_i$ are all equal, i.e., $r_i=a$. Subsequently, applying the integer constraints imply that $r_i$ is either $\ceil{a}$ or $\lfloor a\rfloor$. The structure of the optimal placement vector is then,
\begin{align}
    r_i = 
    \begin{cases}
        \lfloor a\rfloor, &  \# \alpha_1(\ceil{a}-a),\\
        \ceil{a},  &  \# \alpha_1(1+a-\ceil{a}).
    \end{cases}\label{eqn:r_i}
\end{align}
where $\#$ denotes the number of elements of each term. 

Now that we know the structure of the optimal placement vector, next we need to find the optimal arrangement of $\ceil{a}$ and $\lfloor a\rfloor$ terms within the placement vector. This is one of the key differences from the work in \cite{gau_ciss24} where the ordering of the placement vector is inconsequential. Note that to minimize $\tilde{r}(i)$, we need to spread the elements such that they are as uniform as possible within every window of consecutive $i$ terms. This is achieved by hierarchically spreading different sub-blocks of the placement vector as given by Algorithm~\ref{alg:place}.

\begin{algorithm}[t]
\caption{Uniform arrangement of placement vectors}\label{alg:place}
\begin{algorithmic}
\renewcommand{\algorithmicrequire}{\textbf{Input:}}
\Require $\alpha_1$,$\alpha_2$
\State $a=\frac{\alpha_2}{\alpha_1}$, $b_1=\lfloor a \rfloor$, $b_2=\ceil{a}$
\State $c_1=\alpha_1(\ceil{a}-a)$, $c_2=  \alpha_1(1+a-\ceil{a})$
\While{$\min(c_1,c_2)>1$}
    \If {$c_1>c_2$}
        \State $c_1 \gets c_2$, $c_2 \gets c_1$
        \State $b_1 \gets b_2$, $b_2 \gets b_1$
    \EndIf
    \State $c = \frac{c_2}{c_1}$
    \State $b_1 \gets \{b_1,b_2 \times \lfloor c \rfloor\}$ , $b_2 \gets \{b_1,b_2 \times \ceil{c}\}$
    \State $c_1 \gets c_1(\ceil{c}-c)$, $c_2 \gets c_1(1+c-\ceil{c})$
\EndWhile
\end{algorithmic}
\end{algorithm}

In Algorithm~\ref{alg:place}, $b_2 \gets \{b_1,b_2 \times \ceil{c}\}$ implies that the new $b_1$ is an array consisting of one instance of $b_1$ and $\ceil{c}$ instances of $b_2$. As an application of Algorithm~\ref{alg:place}, let us consider a scenario with $\alpha_2 = 41$ and $\alpha_1=11$. One possible placement vector for this case is $\bm{r}=\{3,3,3,4,4,4,4,4,4,4,4\}$ which is a vector of $3$ threes and $8$ fours. Now, we need to uniformly arrange this placement vector. For this purpose, we try to uniformly distribute the $8$ fours among the $3$ threes. This will give us two blocks of $\{3,4,4,4\}$ and one block of $\{3,4,4\}$. Next, we need to spread these sub-blocks as uniformly as possible. Since the minimum of the number of instances of these two blocks is one, the algorithm stops at this stage and returns $r=\{3,4,4,4,3,4,4,4,3,4,4\}$ as the optimal placement vector. The optimal placement vector generated by Algorithm~\ref{alg:place} is always a mixture of placement vectors of the form $r=\{\Theta\}$ and $r=\{\Theta+1\}$, where $\Theta = 3$ in this example.

Note that the above treatment tries to minimize $\e[\Delta_1]$ by scheduling source-$1$ transmissions as uniformly as possible for a fixed $\alpha_1$ and $\alpha_2$. When we uniformly distribute source-$1$ transmissions, we also notice that at the same time, it allows us to uniformly distribute the $S_2$ as well. Hence, the above structure of the placement vector jointly minimizes both $\e[\Delta_1]$ and $\e[\Delta_2]$. This is one of the similarities with the work in \cite{gau_ciss24}. Therefore, what remains is to find the best possible $\alpha_1$ and $\alpha_2$. This can be reduced to finding the best possible rational number for $a$ that minimizes the weighted AoI. 

To find the optimal $a$, we first find the average AoI of the round robin (RR) policy, denoted by $\e[\Delta_{RR}]$, which can be obtained by setting  $\alpha_1 = \alpha_2 = a = 1$ in (\ref{eqn:aoi_1}) and \eqref{eqn:aoi_2},
\begin{align}
    \e[\Delta_{RR}] = \frac{(s_1+s_2)}{2}\left(w_1\frac{(1+p_1)}{(1-p_1)}+w_2\frac{(1+p_2)}{(1-p_2)}\right)+\frac{v_1+v_2}{2(s_1+s_2)}+s_1w_1+s_2w_2.
\end{align}
Since the first term in $\e[\Delta_1]$ is linear in $a$, there will be an $a$ for which the weighted average AoI of source-$1$ will be greater than that of the RR policy. We do not have to consider the values of $a$ beyond this point since then the AoI performance of the RR policy will be superior. Therefore, we only have to consider the values of $a$ for which the term  $w_1\e[\Delta_1]$ is less than the weighted AoI of the RR policy. Let $a_{max}$ be the smallest $a$ such that $w_1\e[\Delta_1]> \e[\Delta_{RR}]$. Similarly, we note that the first term in  $\e[\Delta_2]$ is linear in $\frac{1}{a}$. Therefore, we can find a $a_{min}$ value, which is the largest $a$ such that $w_2\e[\Delta_2]> \e[\Delta_{RR}]$. Hence, the search space of $a$ is bounded by $a_{min}$ and $a_{max}$.

From this point onwards, we represent a two source cyclic schedule  only using the tuple $(\alpha_1,\alpha_2)$ where the optimal placement vector is found using Algorithm~\ref{alg:place}. For a fixed $a$, consider the patterns  $(\alpha_1,\alpha_2)$ and $(k\alpha_1,k\alpha_2)$  where $k\in \mathbb{N}$. If $\bm{r}$ is the optimal placement vector for $(\alpha_1,\alpha_2)$, by simply repeating $\bm{r}$, $k$ times, we can obtain the optimal placement vector for the pattern $(k\alpha_1,k\alpha_2)$ which yields the same average AoI. Therefore, in the bounded region for $a$, we only need to compare rationals in their simplest form. Algorithm~\ref{alg:optim} can be used to find a near-optimal $a$ and $\bm{r}$ which minimize the weighted AoI. In Algorithm~\ref{alg:optim}, $\e[\Delta_i^r]$ represents the AoI of the $i$th source with respect to the placement vector of source-$1$, $\mathbf{ALG\ref{alg:place}}(\tilde{\alpha}_1,\tilde{\alpha}_2)$ is the output of Algorithm~\ref{alg:place} for the selected $\tilde{\alpha}_1$, $\tilde{\alpha}_2$ and $\mathbf{Coprime}(\alpha_1,\alpha_2)$ returns the co-primes of $\alpha_1$ and $\alpha_2$. Even though the average AoI depends on both $a$ and $\alpha_1$, we show in the next section that fixing $\alpha_1$ to a large value and finding $a$ using Algorithm~\ref{alg:optim} is sufficient to be as close as desired to the optimal schedule.

\begin{algorithm}[t]
    \caption{Algorithm to find the near-optimal cyclic pattern}\label{alg:optim}
    \begin{algorithmic}
        \renewcommand{\algorithmicrequire}{\textbf{Input:}}
        \Require $\alpha$ sufficiently large integer.
        \State $\alpha_1= \alpha$, $\alpha_2=\alpha$, $AoI_{min}=\e[\Delta_{RR}]$, $r^*=\{1\}$
        \State $a_{max} = \inf\{a:w_1\e[\Delta_1]>\e[\Delta_{RR}]\}$
        \While{$\frac{\alpha_2}{\alpha_1}<a_{max}$}
            \State $(\tilde{\alpha}_1,\tilde{\alpha}_2) = \mathbf{Coprime}(\alpha_1,\alpha_2)$ 
            \State $r= \mathbf{ALG\ref{alg:place}}(\tilde{\alpha}_1,\tilde{\alpha}_2)$ 
            \State $AoI = w_1\e[\Delta_1^r]+w_2\e[\Delta_2^r]$
            \If {$AoI_{min}> AoI$}
                \State $r^*=r$ , $AoI_{min}=AoI$
            \EndIf
            \State $\alpha_2=\alpha_2+1$
        \EndWhile
        \State $\alpha_1= \alpha$, $\alpha_2=\alpha$
        \State $a_{min} = \sup\{a:w_2\e[\Delta_2]>\e[\Delta_{RR}]\}$
        \While{$\frac{\alpha_2}{\alpha_1}>a_{min}$}
            \State $(\tilde{\alpha}_1,\tilde{\alpha}_2) = \mathbf{Coprime}(\alpha_1,\alpha_2)$  
            \State $r= \mathbf{ALG\ref{alg:place}}(\tilde{\alpha}_1,\tilde{\alpha}_2)$ 
            \State $AoI = w_1\e[\Delta_1^r]+w_2\e[\Delta_2^r]$
            \If {$AoI_{min}> AoI$}
                \State $r^*=r$, $AoI_{min}=AoI$
            \EndIf
            \State $\alpha_1=\alpha_1+1$
        \EndWhile
    \State{\textbf{Output:}} $r^*$
    \end{algorithmic}
\end{algorithm}

\subsection{Proof of Optimality}
In this section, we will prove that for all $\epsilon>0$, the minimum weighted AoI obtained by the solution of Algorithm~\ref{alg:optim} can be made to be within $\epsilon$ of the actual optimum by choosing $\alpha$ in Algorithm~\ref{alg:optim} sufficiently large. Without loss of generality, assume optimal $a^*>1$. Let us analyze the age expression of source-$1$. Define the functions $f(a)$, $g(a)$, and $h(a)$ as follows,
\begin{align}
    f(a)& = \frac{(1+p_1)}{2(1-p_1)}s + \frac{v}{2s}+s_1,\\
    g(a) &= a^2s_2^2\frac{(1-p_1)^2}{2s(1-p_1^{\alpha_1})}\sum_{i=1}^{\alpha_1}i^2p_1^{i-1},\\
    h(a) &= s_2^2\frac{(1-p_1)^2}{2s\alpha_1(1-p_1^{\alpha_1})}\sum_{i=1}^{\alpha_1}\tilde{r}(i)p_1^{i-1}.
\end{align}
Then, $\e[\Delta_1]$ given in \eqref{eqn:aoi_1} is equivalent to,
\begin{align}
\e[\Delta_1] = f(a)-g(a)+h(a).
\end{align}
In Algorithm~\ref{alg:optim}, we fix $\alpha_1$ and increase $\alpha_2$ since $a^*>1$. Since $f(a)$ is a continuous function of $a$, if $a$ is sufficiently close to $a^*$, $f(a) \approx f(a^*)$. For sufficiently large $\alpha_1$, $\sum_{i=1}^{\alpha_1}i^2p_1^{i-1}/(1-p_1^{\alpha_1})$ is approximately a constant. Therefore, when $a$ is sufficiently close to $a^*$, $g(a)\approx g(a^*)$. Consider an $\alpha_0 \in \mathbb{N}$ such that $a_0=\frac{\alpha_0}{\alpha}<a^*<\frac{\alpha_0+1}{\alpha}=a_1$. If $\alpha$ is sufficiently large, for all $a$ such that $a_0<a<a_1$, $\ceil{a}=\ceil{a^*}$ and $\lfloor a \rfloor = \lfloor a^* \rfloor$. Thus, the placement vector for any $a \in [a_0,a_1]$ consists of either $\ceil{a^*}$ or $\lfloor a^* \rfloor$. 

Now, consider $a = (a_0+a_1)/2 = (2\alpha_0+1)/2\alpha$. The average AoI when $(\alpha_1,\alpha_2) = (\alpha,\alpha_0)$ is equal to the average AoI when $(\alpha_1,\alpha_2)= (2\alpha,2\alpha_0)$. The placement vector of $(2\alpha,2\alpha_0+1)$ would only differ in one term with the placement vector of $(2\alpha_0,2\alpha)$, where it is increased by 1.  Therefore, the difference in $\tilde{r}(i)$ of the two patterns will be bounded by $2(\ceil{a^*}+1)i^2$. Therefore, the difference in the average AoI is bounded by $2C(\ceil{a^*}+1)\sum_{i=1}^{\alpha_1}i^2p_1^{i-1}/\alpha_1$, where $C$ is a constant (since $\alpha$ is large, we have made the assumption that $(1-p_1^{\alpha_1}) \approx 1$). Therefore, by approaching $a^*$ in a bisection search, we can show that the AoI difference when $a=a^*$ and when $a$ equals one of the end points $a_0$ or $a_1$, will be bounded by a constant times $1/\alpha$. A similar argument holds for $\e[\Delta_2]$. Therefore, by selecting a large enough $\alpha$, we can make our solution as close as desired to the optimal solution. 

\subsection{Sub-Pattern Refinement}
Algorithm~\ref{alg:optim} provides a means for finding a cyclic schedule as almost as good as the optimal cyclic schedule. However, it may end up generating long placement vectors to be as close as possible to  the optimal. For example, suppose the optimal pattern is known to be $\{2,3,3\}$ where optimal $\alpha_1 = 3$. However, if we set $\alpha=101$ in Algorithm~\ref{alg:optim}, it would never output the pattern $\{2,3,3\}$ since $\alpha$ is not divisible by $3$. Instead, it would output a pattern like $\{2,3,3,3,2,3,3,2,3,3,\dots,2,3,3\}$, where the sub-pattern $\{2,3,3,3\}$ is followed by multiple repetitions of the sub-pattern $\{2,3,3\}$ so that the average AoI would be as close as possible to the optimal. Thus, we propose to further check if the sub-patterns existing within the output of Algorithm~\ref{alg:optim}, could be a better solution. Algorithm~\ref{alg:refine} provides a method to further  refinea a cyclic schedule using these sub-patterns to produce a better and a simpler schedule.

\begin{algorithm}[t]
    \caption{Pseudo-code to refine the schedule using sub-patterns}\label{alg:refine}
    \begin{algorithmic}
        \State{\textbf{Step 1:}} Set $\alpha$ sufficient large and find the near optimal $a^*$ and $\alpha_1^*$ from Algorithm \ref{alg:optim}
        \State{\textbf{Step 2:}} Set $\alpha_1=\alpha_1^*$ and $\alpha_2=a^*\alpha_1^*$ in Algorithm \ref{alg:place}. Let $SB$ be the set of all sub-blocks constructed across all stages in Algorithm \ref{alg:place}.
        \State {\textbf{Step 3:}} Find the sub-block in $SB$ with the lowest weighted AoI.
    \end{algorithmic}
\end{algorithm}

\section{Multi-Source Cyclic Schedulers ($N>2$)}\label{sec:multi}
We have outlined the structure followed by a near optimal scheduler for two heterogeneous sources in Section~\ref{sec:2source}. Even though the algorithms used to construct the two source scheduler can be proven to be optimal, they cannot be easily extended to handle a larger number of sources ($N>2)$. In this section, we propose an algorithm to construct cyclic schedulers for a large number of sources. The algorithms  used here are derived using the structure of the age expression in \eqref{exp_scov} presented in our MGF formulation.

\subsection{Scalable Weighted AoI Minimization Scheduler (SAMS)}
Let $\tau_n$, $0 < \tau_n < 1$, $n \in \{1,\ldots,N\}$, denote the link utilization of source-$n$, i.e., long-term fraction of time that the wireless link is occupied with the transmission of packets (successful or unsuccessful) from source-$n$. Since we focus on work-conserving servers, we have $\sum_{n=1}^N \tau_n =1$. Moreover, let the random variable $\tilde{S}_n^u$ (steady-state random variable associated with the random process $\tilde{S}_{n,k}^u$ as $k \rightarrow \infty$) have mean $\tilde{s}_n^u =\mathbb{E} [\tilde{S}_n^u]$. Then, $\tau_n$ takes the following form,
\begin{align}
    \tau_n = \frac{s_n + s^u_n}{s_n + \tilde{s}_n}. \label{nail41}
\end{align}
For every successful source-$n$ transmission, we must have on  average $\frac{p_n}{u_n}$ unsuccessful transmissions from the same source. Therefore, we have $s^u_n = \frac{p_n}{u_n} s_n$, from which we obtain $\tilde{s}_n$ in terms of $\tau_n$ as,
\begin{align}
  \tilde{s}_n & = \frac{s_n}{u_n \tau_n} - s_n. \label{nail43}  
\end{align} 
By substituting \eqref{nail43} into \eqref{exp_scov}, we obtain,
\begin{align}
2 w_n \mathbb{E}[\Delta_n]  &=  \underbrace{w_n s_n u_n (c_n + \tilde{c}_n)}_{a_n} \tau_n + \underbrace{\frac{w_n s_n (1+\tilde{c}_n)}{u_n}}_{b_n} \frac{1}{\tau_n} + 2 w_n s_n (1 - \tilde{c}_n), \label{nail51}\\
 &= \frac{(u_n\tau_n-1)^2}{u_n\tau_n}w_ns_n\tilde{c}_n+w_ns_n\left(c_nu_n\tau_n+\frac{1}{u_n\tau_n}+2\right). \label{nail51_b}
\end{align}
From \eqref{nail51}, we can obtain the weighted AoI as,
\begin{align}
    \e[\Delta]=\sum_{n=1}^N \frac{a_n}{2}\tau_n+\frac{b_n}{2\tau_n}+ \sum_{n=1}^N w_ns_n(1-\tilde{c}_n).
\end{align}

The source utilization variables $\tau_n$ affect directly the parameters $\tilde{s}_n$ but not the scov parameters $\tilde{c}_n$ in \eqref{exp_scov}. In fact, $\tilde{c}_n$s are complex functions of the scheduler itself and therefore it is difficult to pose a joint optimization problem involving both $\tilde{s}_n$ and $\tilde{c}_n$ for $1 \leq n \leq N$. From \eqref{nail51_b}, we note that, for a fixed choice of $\tau_n$ values, the best choice for the minimization of weighted AoI would be to schedule the sources in such a way that $\tilde{c}_n$ parameters are as close to zero as possible, i.e., close to periodic transmissions of successful source-$n$ packets. However, this may not always be possible even in the case of deterministic service times. Even if source-$n$ transmissions are periodic, the time interval between two consecutive successful source-$n$ transmissions may not necessarily be deterministic, i.e., $\tilde{c}_n$ will be non-zero, which stems from non-zero packet drop probability for source-$n$. Moreover, it may not be possible for source-$n$ transmissions to be periodic due to constraints from other sources. When the service times are random, this situation becomes  even more challenging. Therefore, to overcome this issue, we first find the source utilization parameters for fixed a set of $\tilde{c}_n$ values. This leads us to the following optimization problem,
\begin{mini}
    {\tau_n \geq 0}{\sum_{n=1}^{N} \left( a_n \tau_n +  b_n \frac{1}{\tau_n} \right) }
    {\label{Optimization1}}
    {}
    \addConstraint{ \sum_{n=1}^{N} \tau_n}{= 1},
\end{mini}
where the coefficients $a_n$ and $b_n$ are non-negative. The optimization problem \eqref{Optimization1} is convex and the KKT conditions result in the following condition, 
\begin{align}
    a_n - \frac{b_n}{\tau_n^2} & = a_m - \frac{b_m}{\tau_m^2} , \qquad 1 \leq n,m \leq N,  
\end{align}
which together with the normalization constraint in \eqref{Optimization1} yields the following non-linear fixed point equation in the single unknown $x \in \mathbb{R}$,
\begin{align}
    f_p(x) & = \sum_{n=1}^N \sqrt{\frac{b_n}{a_n-x}} -1 = 0. \label{fixedpoint}
\end{align}
We note that the function $f_p(x) \in (-1,\infty)$ is a monotonically increasing function of $x \in (-\infty,\min\limits_{1 \leq n \leq N} a_n)$,
and therefore, has a unique solution denoted by $x^*$ (can be found by bisection search) from which one can find the optimum coefficients $\tau_n^*$, $1\leq n \leq N$, from, 
\begin{align}
    \tau_n^* & = \sqrt{\frac{b_n}{a_n-x^*}}. 
\end{align}
Let $f_n$ denote the long-term frequency of packet transmissions from source-$n$. Since $\tau_n \propto f_n s_n$ (by definition), the optimum transmission frequency, denoted by $f_n^*$, $1 \leq n \leq N$, is given by the following closed-form expression,
\begin{align}
    f_n^* & = {\frac{\tau_n^*}{s_n}} \left( {\sum\limits_{m=1}^N \frac{\tau_m^*}{s_m}} \right)^{-1}. \label{eqn:optimum_frequencyAoI}
\end{align}
\subsubsection{Cyclic Schedule Construction}
Next, we present how to construct a well-performing cyclic scheduler based on the transmission frequencies computed above. As the next step for cyclic scheduling, we first need to write $f_n^* \approx K_n/K$ for integer $K_n$, $K$  where $K_n$ would represent the number of occurrences of source-$n$ within the pattern $P$. For this, we propose Algorithm~\ref{alg:General} in terms of the algorithm parameter $\varepsilon \geq 0$. The case of $\varepsilon =0$ ensures that $K_n \geq 1, \; \forall n,$ and has a small pattern size $K$. When $\varepsilon$ increases, obviously the approximation $K_n/K$ of the optimum frequency $f_n^*$ improves but at the expense of increased $K$, and hence, increased storage requirements. The use of $\varepsilon>0$ was motivated by the fact that NOTS produced optimal patterns that were a mixture of two sub-patterns. Therefore, by allowing the minimum number of source scheduling instances to be greater than 1, we make our policy more flexible and closer to  NOTS when $N=2$. The computational complexity of Algorithm~\ref{alg:General} is $O( N\log N)$ due to the required sorting.

\begin{algorithm}[t]
    \caption{Pseudo-code for obtaining $K_n$ and $K$}
    \begin{algorithmic}
    \renewcommand{\algorithmicrequire}{\textbf{Input:}}
    \renewcommand{\algorithmicensure}{\textbf{Output:}}
        \Require $f_n^*, 1 \leq n \leq N$, $\varepsilon \geq 0$;
        \Ensure $K_n,K$;
        \State {\bf Step 1:} Find $f_{min}^* =\min_n f_n^*$;
        \State {\bf Step 2:} Set $K = \ceil{\frac{(1+\varepsilon)}{f_{min}}}$;
        \State {\bf Step 3:} Find $R = \sum_{n=1}^N \lfloor Kf_n\rfloor$;
        \State {\bf Step 4:} Sort all the sources in descending order according to the fractional part of $K f_n$ and set $K_n = \ceil{K f_n}$ for the top $K-R$ sources in this ordered list, otherwise set $K_n = \lfloor Kf_n \rfloor$.    
    \end{algorithmic}
    \label{alg:General}
\end{algorithm}

Next, we need a packet spreading algorithm which generates a transmission pattern in which source-$n$ appears $K_n$ times with pattern size $K=\sum_{n=1}^N K_n$, and all these appearances are as evenly spread as possible throughout the pattern. This problem with many variations had been studied in the context of \emph{internet scheduling} where the goal was to share the link bandwidth fairly among multiple flows carrying fixed-size or variable-size packets; see \cite{shreedhar_varghese_sigcomm95} and the references therein for a collection of research papers on fair link sharing.

The packet spreading algorithm we propose to use in the present paper is based on the \emph{deficit round robin} (DRR) algorithm proposed in \cite{shreedhar_varghese_sigcomm95} which has been successfully used in commercial routers due to its low computational complexity. DRR consists of rounds at each of which the deficit counters of each flow are incremented by the product of the so-called quantum and the weight of the flow. Subsequently, all the head-of-line packets waiting in the queue of each flow whose total packet size in bytes does not exceed the corresponding deficit counter, are served. The number of bytes that are served for a flow is then subtracted from the corresponding deficit counter, leaving a so-called deficit. In this way, multiple flows can be served in the same round. In our proposed spreading algorithm, our goal is to find a cyclic scheduler that spreads out source-$n$ transmissions as much as possible while source-$n$ appears $K_n$ times in the pattern. Initially, all deficit counters, denoted by $B_n(t)$, $1 \leq n \leq N$, are set to zero. We modify the original DRR algorithm by allowing the value of the quantum change between rounds, which is slightly different than the original DRR scheduler \cite{shreedhar_varghese_sigcomm95}. Particularly, the quantum is chosen to ensure that one source is guaranteed to be inserted into the pattern at a given round while leaving zero deficit. The pseudo-code for the proposed packet spreading algorithm is given in Algorithm~\ref{alg:Spreading} with computational complexity ${O}(NK)$.

\begin{algorithm}[t]
    \caption{Algorithm for constructing the pattern $P$}
    \begin{algorithmic}
    \renewcommand{\algorithmicrequire}{\textbf{Input:}}
    \renewcommand{\algorithmicensure}{\textbf{Output:}}
        \Require $N \geq 2$, $K_n, n=1,\ldots,N$, $K=\sum_{n=1}^N K_n$;
        \Ensure Pattern $P$ of size $K$;
        \State $B_n(t) \gets 0$; $\;$ ($B_n(t)$: deficit counter for source-$n$)
        \For {$k=0$ \textbf{to} $K-1$}
        \State $m \gets \arg \min\limits_{ 1 \leq n \leq N} \frac{(1-B_n(t))K}{K_n}$;  (ties broken randomly)
        \State $Q \gets \frac{(1-B_m(t))K}{K_m}$; ($Q$: quantum)
        \For {$n=1$ \textbf{to} $N$}
        \State $B_n(t) = B_n(t) + Q \frac{K_n}{K}$; (update deficit counters)
        \EndFor
        \State $P(k) \gets m$; (insert source-$m$ in $P$)
        \State $B_m(t) \gets 0$;
        \EndFor
    \end{algorithmic}
    \label{alg:Spreading}
\end{algorithm}

\subsubsection{SAMS Construction}
Note that the optimal frequencies given by \eqref{eqn:optimum_frequencyAoI} depend on the value we choose for $\tilde{c}_n$. However, $\tilde{c}_n$ itself is an artifact of the constructed scheduler. Hence, there is a circular dependency between $\tilde{c}_n$ and the constructed scheduler. Thus, we propose the following  fixed point iterative method for the construction of SAMS. As the first step in this construction, we need to find a good initial approximation for $\tilde{c}_n$. For this purpose, we assume that the scheduler is able to transmit source-$n$ packets with a period of $T_n = \frac{s_n}{\tau_n}$ which will ensure a link utilization of $\tau_n$. With this assumption in hand, $\tilde{S}_n$ behaves as $\tilde{S}_n = T_n X_n - S_n$, where $X_n$ is a geometric random variable with parameter $u_n$. Recalling that a geometrically distributed random variable with parameter $u_n$ has mean $\frac{1}{u_n}$ and variance $\frac{1-u_n}{u_n^2}$, we have,
\begin{align}
 \mathbb{E} [\tilde{S}_n]  = \frac{T_n}{u_n} - s_n, \qquad \text{Var} [\tilde{S}_n]  =  \frac{T_n^2 p_n}{u_n^2} + v_n. 
\end{align}
In a large-scale status update system, $T_n \gg s_n$ and $T_n^2 \gg v_n$. Hence, we propose to approximate $\tilde{c}_n = 
\frac{\text{Var} [\tilde{S}_n] }{ \mathbb{E} [\tilde{S}_n]^2}$ by $p_n$ which does not depend on $\tau_n$. Therefore, we set our initial approximation of $\tilde{c}_n$ as $\tilde{c}_n^{(0)}=p_n$.

Starting with $\tilde{c}_n^{(0)}$, we run our fixed point iteration algorithm for $L$ iterations. In iteration $\ell$, $1 \leq \ell\leq L$, for a given value of $\tilde{c}_n^{(\ell-1)}$, we obtain the source-$n$ frequency $f_n^{*(\ell)}$ according to \eqref{eqn:optimum_frequencyAoI}. Then, we choose a $\varepsilon \geq 0$ out of a given set $\mathcal{E}$ of $\varepsilon$ values and employ Algorithms~\ref{alg:General} and \ref{alg:Spreading} to obtain a pattern $P^{(\ell,\varepsilon)}$ in iteration $\ell$ whose corresponding weighted AoI $\mathbb{E}[\Delta^{(\ell,\varepsilon)}]$ and $\tilde{c}_n{^{(\ell,\varepsilon)}}$ values can be calculated based on the procedure described in Section~\ref{sec:mgf_analysis}. We repeat the process for each value $\varepsilon$ in the set $\mathcal{E}$ and choose the particular value of $\varepsilon$, denoted by $\varepsilon'$, resulting in a pattern $P^{(\ell)}=P^{(\ell,\varepsilon')}$ with the minimum weighted AoI in the iteration $\ell$. Then, the parameter $\tilde{c}_n{^{(\ell+1)}}=\tilde{c}_n{^{(\ell,\varepsilon')}}$ is fed as input to iteration $\ell+1$, and this process is repeated for $L$ iterations. The pattern $P^{(\ell)}$ which generates the lowest weighted AoI among all the iterations $1 \leq \ell \leq L$ is then output by our proposed algorithm, called SAMS (scalable weighted AoI minimizing scheduler), which is a function of the algorithm parameters $\mathcal E$ and $L$. Hence, the proposed method SAMS should not be viewed as a vector fixed-point iteration but rather a search algorithm along the fixed-point iterations. For a given $\varepsilon$, the computational complexity of one SAMS iteration is ${O}(NK)$ due to the fact that Algorithm~\ref{alg:Spreading} dominates the overall execution time, where $K$ is the size of the pattern produced by SAMS. Algorithm~\ref{alg:SAMS} presents the pseudo-code for SAMS construction. 

\begin{algorithm}[t]
    \caption{Psuedo code for SAMS construction}
    \begin{algorithmic}
    \renewcommand{\algorithmicrequire}{\textbf{Input:}}
    \renewcommand{\algorithmicensure}{\textbf{Output:}}
        \Require $\mathcal{E}$, $L$;
        \Ensure SAMS;
        \State $\tilde{c}_n^{(0)}=p_n \; \forall \; n$
        \For {$\ell=0$ \textbf{to} $L-1$}
        \State $f_n^{*(l)} \Leftarrow \tilde{c}_n^{(l)} \; \forall \; n$  (compute optimal frequencies from $\tilde{c}_n^{(l)}$)
            \For {$\varepsilon \in \mathcal{E}$}
                \State $\{K_n^\varepsilon\}_{n=1}^N=\mathbf{ALG\ref{alg:General}}(\{f_n^{*(l)}\}_{n=1}^N,\varepsilon)$ (obtain the $K_n$ values from Algorithm \ref{alg:General})
                \State $P^{(\ell,\varepsilon)}$ = $\mathbf{ALG\ref{alg:Spreading}}(\{K_n^\varepsilon\}_{n=1}^N)$ (construct the pattern using Algorithm \ref{alg:Spreading})
                \State $\mathbb{E}[\Delta^{(\ell,\varepsilon)}],\{\tilde{c}_n{^{(\ell,\varepsilon)}}\}_{n=1}^N \Leftarrow P^{(\ell,\varepsilon)} $ (obtain the weighted AoI and the  $\tilde{c}_n$ for pattern $P^{(\ell,\varepsilon)}$)
            \EndFor
            \State $\varepsilon' = \arg\underset{\varepsilon}{\min} \{ \mathbb{E}[\Delta^{(\ell,\varepsilon)}]\}$
            \State  $\tilde{c}_n{^{(\ell+1)}}=\tilde{c}_n{^{(\ell,\varepsilon')}}$, $P^{(l)} = P^{(\ell,\varepsilon')}$, $\mathbb{E}[\Delta^{(\ell)}]=\mathbb{E}[\Delta^{(\ell,\varepsilon')}]$
        \EndFor
        \State $\ell' = \arg\underset{\ell}{\min} \{\mathbb{E}[\Delta^{(\ell)}]\} $
        \State SAMS $ = P^{(\ell')}$
    \end{algorithmic}
    \label{alg:SAMS}
\end{algorithm}

\subsection{Characteristics of the Packet Spreading Algorithm} \label{sec:drr_prop}
In this section, we present a few properties of the packet spreading algorithm presented in Algorithm~\ref{alg:Spreading} that ensures the generation of a cyclic schedule which adheres to the number of occurrences $\{K_n\}_{n=1}^{N}$ of each source given by Algorithm~\ref{alg:General}. In particular, we would like to show that the output pattern would contain exactly $K_n$ occurrences of source-$n$ and all the deficit counters would be zero at the end of $K$ iterations.

Let $B_n^k$ denote the value of the deficit counter for the $n$th source in the $k$th iteration of Algorithm \ref{alg:Spreading}. Let $Q_n^k$ denote the quantum allocated for source-$n$ in the $k$th iteration, where $Q_n^k = \frac{(1-B_n^k)K}{K_n}$. Since the initial deficit counters are all set to zero, $Q_n^0= \frac{1}{K_n}$. In each iteration, the source with the smallest quantum is selected as the source to be inserted into the pattern $P$. Then, the deficit counter of the selected source is set to zero again and the deficit counters of all the other sources are updated by the equation $B_n^{k+1} = B_n^k+ Q\frac{K_n}{K}$, where $Q=\min_n Q_n^k$. Based on the above steps, Lemma~\ref{lem:q_update} gives the update equations for the quantum allocated for each source.

\begin{lemma}\label{lem:q_update}
    The quantum $Q_n^k$ allocated for source-$n$ is updated according to the equation,
    \begin{align}
        Q_n^{k+1}=
        \begin{cases}
            Q_n^k-\min_nQ_n^k, & \quad n \neq \arg \underset{j}{\min} \;Q_j^k, \\
            \frac{K}{K_n}, & \quad n = \arg\underset{j}{\min}\;Q_j^k.
        \end{cases}
    \end{align}
\end{lemma}

\begin{Proof}
    If  $n \neq \arg \underset{j}{\min} \;Q_j^k $, then, the quantum in the next iteration is given by,
    \begin{align}
        Q_n^{k+1}= \frac{(1-B_n^{k+1})K}{K_n}= \frac{(1-B_n^k-Q\frac{K_n}{K})K}{K_n}=\frac{(1-B_n^k)K}{K_n}-Q = Q_n^k-\min_nQ_n^k.
    \end{align}
    If $n = \arg\underset{j}{\min}\;Q_j^k$, then, $B_n^{k+1}=0$. Therefore, $Q_n^{k+1}=\frac{K}{K_n}$.
\end{Proof}

Since all $Q_n^k$s have a scaling factor of $K$, we can safely normalize all the $Q_n^k$ by $K$. Let $\tilde{Q}_n^k$ represent the normalized $Q_n^k$ values. Then, the source to be inserted in each iteration is given by $\arg \underset{j}{\min} \;\tilde{Q}_j^k$, and its update equations are given by,
\begin{align}
        \tilde{Q}_n^{k+1}=
        \begin{cases}
            \tilde Q_n^k-\min_n\tilde Q_n^k, & \quad n \neq \arg \underset{j}{\min} \;\tilde Q_j^k, \\
            \frac{1}{K_n}, & \quad n = \arg\underset{j}{\min}\; \tilde Q_j^k.
        \end{cases}
\end{align}
    
We will first prove the desired properties for $N=2$, and later show how it can be extended for $N>2$. Without loss of generality, assume $K_1\geq K_2 \geq \dots \geq K_N$. Let $x_n^k$ represent the number of source-$n$ instances that had been inserted into the pattern after the $k$th iteration.

\begin{property}\label{prop:1}
    For $N=2$, $\tilde Q_{N-i+1}^{k+1}=\frac{x^k_{N-i+1}+1}{K_{N-i+1}}-\frac{x^k_i}{K_i}$, where $i$ is the source that was selected in the $k$th iteration.
\end{property}

\begin{Proof}
Initially, the pattern $P$ is empty with $\tilde Q^0_1=\frac{1}{K_1}$ and $\tilde Q^0_2 =\frac{1}{K_2}$. Since $K_1\geq K_2$, $\tilde Q^0_1\leq \tilde Q^0_2$. Therefore, we would insert a source-$1$ scheduling instance to the pattern $P$. Updating the normalized quantums would yield that $\tilde Q^1_1=\frac{1}{K_1}$ and $\tilde Q^1_2=\frac{1}{K_2}-\frac{1}{K_1}$. Note that Property~\ref{prop:1} is true for $k=0$. We will now prove that the result will hold for $k=m+1$ assuming that it holds for $k=m$. For brevity, let $j=N-i+1$, which is the source that was not selected in the $m$th iteration. Based on our assumption $\tilde Q_{j}^{m+1}=\frac{x^m_{j}+1}{K_{j}}-\frac{x^m_i}{K_i}$, where $i$ is the source that was selected in the $m$th iteration. Since $i$ was selected in the $m$th iteration, $\tilde Q_i^{m+1}=\frac{1}{K_i}$. There are two possibilities to consider. If source-$i$ was selected again in the $(m+1)$th iteration, then $x^{m+1}_{j}=x^m_{j}$ and $x^{m+1}_i=x^m_i+1$. Therefore, $\tilde Q_{j}^{m+2}=\frac{x^m_{j}+1}{K_{j}}-\frac{x^m_i+1}{K_i}= \frac{x^{m+1}_{j}+1}{K_{j}}-\frac{x^{m+1}_i}{K_i}$. This satisfies Property \ref{prop:1}. If source-$j$ was selected in the  $(m+1)$th iteration, then $x^{m+1}_{j}=x^m_{j}+1$ and $x^{m+1}_i=x^m_i$. Therefore, $Q_i^{m+2}=\frac{x^m_i+1}{K_i}-\frac{x^m_{j}+1}{K_{j}}= \frac{x^{m+1}_i+1}{K_i}-\frac{x^{m+1}_{j}}{K_{j}}$. This too satisfies Property \ref{prop:1}. Therefore by induction, Property~\ref{prop:1} holds for all $k\geq0$.
\end{Proof}

\begin{property} \label{prop:2}
    For $N=2$, $x_n^{K-1}=K_n$.
\end{property}

\begin{Proof}
    Since $\tilde Q^0_1\leq \tilde Q^0_2$, we would insert source-$1$ scheduling instances into the pattern until $\tilde Q^k_2 < \frac{1}{K_1}$. Suppose we insert a source-$2$ scheduling instance for the first time in the $m_1$th iteration. Then, $\tilde Q^{m_1}_2 = \frac{1}{K_2}-\frac{m_1}{K_1} <\tilde Q_1^{m_1}=\frac{1}{K_1}$. Updating the normalized quantums yields, $\tilde Q_2^{m_1+1} = \frac{1}{K_2}$ and $\tilde Q_1^{m_1+1}= \frac{m_1+1}{K_1}-\frac{1}{K_2}$. Here, $m_1$ would correspond to the smallest positive integer such that $\frac{m_1+1}{K_1}-\frac{1}{K_2}>0$. This implies that $\frac{m_1}{K_1}-\frac{1}{K_2}<0$ and $\tilde Q^{m_1+1}_1<\frac{1}{K_1}\leq \frac{1}{K_2}=\tilde Q^{m_1+1}_2$. Therefore, a source-$1$ instance would be inserted to the pattern in the $(m_1+1)$th iteration. Updating the normalized quantums again would yield that $\tilde Q_1^{m_1+2}=\frac{1}{K_1}$ and $\tilde Q_2^{m_1+2}=\frac{2}{K_2}-\frac{m_1+1}{K_1}$. Then, again we will be inserting source-$1$ scheduling instances  into the pattern until $\frac{2}{K_2}-\frac{m_2}{K_1}<\frac{1}{K_1}$. Here, $m_2+1$ is the iteration at which we insert the second instance for source-$2$ into the pattern where $m_2$  corresponds to the smallest positive integer such that $\frac{m_2+1}{K_1}-\frac{2}{K_2}>0$. Moreover, $m_2$ also corresponds to the number of source-$1$ instances inserted into the pattern. Extending this argument $K_2$ times, we get that $K_2$th scheduling instance of source-$2$ will be inserted into the pattern when $\frac{K_2}{K_2}-\frac{m_{K_2}}{K_1}<\frac{1}{K_1}$, where $m_{K_2}$ is the smallest positive integer that satisfies the inequality and $m_{K_2}+K_2-1$ is the iteration at which the inequality is satisfied. This gives us that $m_{K_2}=K_1$, and therefore, the iteration where we insert the $K_2$th scheduling instance of source-$2$ is $K_1+K_2-1=K-1$, proving the desired result.
\end{Proof}

\begin{property}\label{prop:3}
    For $N>1$, $\tilde Q_j^{k+1}=\frac{x^k_j+1}{K_j}-\frac{x^k_i}{K_i}$, where $i$ is the source that was selected in the $k$th iteration and $j\neq i$.
\end{property}

Property~\ref{prop:3} is proved using an induction argument similar to Property~\ref{prop:1}.

\begin{property}
    For $N>1$, $x_n^{K-1}=K_n$ and $\tilde Q_n^K = \frac{1}{K_n}$. Therefore, $B_n^K = 0$.
\end{property}

\begin{Proof}
    Consider source-$j$ and source-$N$ together. Note that, whenever we select another source other than source-$j$ or source-$N$ to be inserted into the pattern, we will be subtracting the same constant from both $\tilde Q_j^k$ and $\tilde Q_N^k$. Therefore, the insertion of source-$j$ relative to source-$N$ insertions would follow the same pattern when the other sources are considered to be absent. Therefore, from Property~\ref{prop:2}, by the time we insert the $K_N$th scheduling instance of source-$N$ into the pattern we would have inserted $K_j$ instances of source-$j$ into the pattern. This is true for all $1\leq j<N$. Therefore, the $K_N$th instance of source-$N$ would be inserted into the pattern in the iteration $K-1$, and thus $x_j^{K-1}=K_j$. Since we selected source-$N$ in the $(K-1)$th iteration, we have $\tilde Q^K_N=\frac{1}{K_N}$. From Property~\ref{prop:3}, we have that $\tilde Q_j^{K}=\frac{x^{K-1}_j+1}{K_j}-\frac{x^{K-1}_N}{K_N}=\frac{K_j+1}{K_j}-\frac{K_N}{K_N}=\frac{1}{K_j}$. This implies that $B_n^K=0$.
\end{Proof}

\subsection{Grouped Packet Spreading Algorithm}
In general, Algorithm~\ref{alg:Spreading} which is used for packet spreading works well for most input vector of the required source occurrences $\{K_n\}_{n=1}^N$. However, we note that for certain pathological input vectors, its output pattern is not exactly uniform. For example, consider an input vector with $5$ sources and the vector of $K_n$ values are $\{8,1,1,1,1\}$. In this particular scenario, the output pattern of the algorithm would be $P=[1,1,1,1,1,1,1,1,2,3,4,5]$. This pattern can be further improved if we can spread the source-$1$ occurrences uniformly among the other source instances leading to a pattern like $P=[1,1,2,1,1,3,1,1,4,1,1,5]$. Therefore, to further improve our packet spreading algorithm we propose the grouped packet spreading algorithm given in Algorithm~\ref{alg:msDRR}.

In Algorithm~\ref{alg:msDRR}, \textbf{GRP}$(V)$ is a function which takes in a vector of the number of source (group) occurrences, groups together the sources (groups) with the lowest identical number of occurrences and outputs the set of groups $G_i$ along with a vector representing the total number of elements within each group $V_i$. \textbf{RRA}$(P,G_j)$ takes in an input set of groups $G_j$ and a pattern $P$, and allocates the elements present in the $l$th group of $G_j$ in a round robin fashion at locations in the pattern $P$ reserved for the $l$th group of $G_j$. 

\begin{algorithm}[t]
    \caption{Multi-stage packet spreading algorithm}
    \begin{algorithmic}
    \renewcommand{\algorithmicrequire}{\textbf{Input:}}
    \renewcommand{\algorithmicensure}{\textbf{Output:}}
        \Require $N \geq 2$, $K_n, n=1,\ldots,N$, $K=\sum_{n=1}^N K_n$;
        \State $V_0 = [K_1,K_2, \dots ,K_N]$ , $G_0= [1,2, \dots, N]$, $i = 1$
        \State $G_1,V_1 = \textbf{GRP}(V_0)$ 
        \While{ $V_i \neq V_{i-1} $}
            \State $G_{i+1},V_{i+1} = \textbf{GRP}(V_i)$ 
            \State $i \gets i +1$
        \EndWhile
        \State $P = \textbf{ALG\ref{alg:Spreading}}(V_{i-1})$
        \For{$j=i-1$  \textbf{to} $1$}
            \State $P=\textbf{RRA}(P,G_j)$
        \EndFor
    \end{algorithmic}
    \label{alg:msDRR}
\end{algorithm}

The general idea behind this algorithm is to group together sources with the same number of occurrences together and treat them as a single source. Once grouped, Algorithm~\ref{alg:Spreading} can be used to find the pattern for this grouped source problem. Once this pattern is found, for every instance a particular group appears in pattern, we assign the sources in that particular group in a round robin manner. This allows us to generate more uniformly spread patterns. When the groups are formed, there may be instances where the number of occurrences of two groups are also identical. Therefore, this grouping procedure must be done several times before applying Algorithm~\ref{alg:Spreading}. This grouped packet spreading algorithm helps further minimize the weighted AoI, but it is at the expense of increased complexity.

As an application of the grouped packet spreading algorithm, let us consider the following input vector of $K_n$ values, $V_0=[16, 2, 1, 1, 2]$. Using $V_0$, the \textbf{GRP} function outputs $G_1=\{[1], [2], [5], [3,4]\}$ and $V_1=[16, 2, 2, 2]$. Here, $G_1$ contains $4$ groups where each group contains a vector representing the positions of their corresponding elements within $V_0$. Note that in $G_1$, only the element $1$ in $V_0$ has been grouped together. Now, if we apply another round of grouping using $V_1$, we get $G_2 =\{[1], [2, 3, 4]\}$ and $V_2=[16, 6]$. At this stage, grouping further would not yield a different vector for $V_3$, when compared to $V_2$. Therefore, we stop the grouping and apply Algorithm~\ref{alg:Spreading} to vector $V_2$. This will give us the following pattern $P=[1, 1, 2, 1, 1, 1, 2, 1, 1, 1, 2, 1, 1, 2, 1, 1, 1, 2, 1, 1, 1, 2]$. Now, for each group in $G_2$, we allocate the corresponding elements in a round robin fashion. This would not change any elements in $P$ with value $1$, but for elements with value $2$, it would assign $[2, 3, 4]$ in a round robin manner. This gives us the updated pattern as $P=[1, 1, 2, 1, 1, 1, 3, 1, 1, 1, 4, 1, 1, 2, 1, 1, 1, 3, 1, 1, 1, 4]$. Next, we do another round robin allocation to this new pattern based on the groups in $G_1$. This gives the final pattern as $P=[1, 1, 2, 1, 1, 1, 5, 1, 1, 1, 3, 1, 1, 2, 1, 1, 1, 5, 1, 1, 1, 4]$.
 
\section{Alternative Age-Agnostic Scheduling Algorithms}\label{sec:other}
In this section, we describe the modifications we have made on two existing age-agnosting scheduling schemes, namely, the insertion search (IS) and probabilistic GAW (P-GAW) scheduling algorithms, that were first proposed and studied in \cite{gau_ciss24}, but for the case when packet errors were absent. We will use them as baseline algorithms (in addition to Eywa \cite{eywa}) to compare them against NOTS for the two-sources case and SAMS for general number of sources.

\subsection{IS Algorithm}
In this section, we modify and extend the IS algorithm which was introduced in \cite{gau_ciss24} and extend it to accommodate packet drops in the channel. IS is an iterative approach for the construction of a cyclic scheduler where it starts off with an RR pattern and iteratively expands this pattern. Suppose the current pattern and pattern size in the $i$th iteration of the IS algorithm are $P^{(i)}$ and $K^{(i)}$, respectively. Then, in each iteration, the IS algorithm selects a source from among the $N$ sources and a location from among the $K^{(i)}$ possible locations, where a new scheduling instance of the selected source can be inserted into the current pattern $P^{(i)}$, to construct a total of $NK^{(i)}$ patterns. The pattern which results in the lowest weighted AoI is then selected as the starting pattern for the next iteration, $P^{(i+1)}$. This procedure is repeated for $I$ iterations and the pattern that resulted in the lowest weighted AoI is selected from the patterns $P^{(i)}$ for $i = 1$ to $I$. One difference when compared to the IS algorithm presented in \cite{gau_ciss24}, is that we do not terminate the algorithm when the pattern in the current iteration has a higher weighted AoI than the pattern in the previous iteration. Instead, we continue for $I$ iterations and find the best possible pattern. The computational complexity of the IS algorithm is $O(N^2I^3)$ and its pseudo-code is presented in Algorithm~\ref{alg:IS}, where $\mathbf{INS}(P,n,k)$ is a function which takes in a pattern $P$ and inserts a new source-$n$ scheduling instance after the $k$th element in the pattern $P$ and outputs this new pattern. Table~\ref{tab:comp} compares the worst case computational complexities of IS along with the other main multi-source scheduling schemes ($N>2$) considered in this work.

\begin{algorithm}[t]
    \caption{Pseudo-code for IS algorithm}
    \begin{algorithmic}
    \renewcommand{\algorithmicrequire}{\textbf{Input:}}
    \renewcommand{\algorithmicensure}{\textbf{Output:}}
        \Require $\{p_n,s_n,c_n\}$ for $1\leq n\leq N$,$I$;
        \Ensure $P^*_{IS}$;
         \State $P^{(N)} = [1,2,\dots,N]$
         \State $\mathbb{E}[\Delta^{(N)}] \Leftarrow P^{(N)}$ (obtain the weighted AoI $\mathbb{E}[\Delta^{(N)}]$ for pattern $P^{(N)}$)
         \For {$i=N+1$ \textbf{to} $I$}
            \For {$k=1$ \textbf{to} $i$}
                \For {$n=1$ \textbf{to} $N$}
                    \State $P_{n,k}^{(i)} = \mathbf{INS}(P^{(i)},n,k)$
                    \State $\mathbb{E}[\Delta_{n,k}^{(i)}] \Leftarrow P_{n,k}^{(i)}$ (obtain the weighted AoI $\mathbb{E}[\Delta_{n,k}^{(i)}]$ for pattern $P_{n,k}^{(i)}$)
                \EndFor
            \EndFor
            \State $n',k' =  \arg\underset{n,k}{\min}\mathbb{E}[\Delta_{n,k}^{(i)}]$
            \State $P^{(i+1)} = P_{n',k'}^{(i)}$
            \State $\mathbb{E}[\Delta^{(i)}] = \mathbb{E}[\Delta_{n',k'}^{(i)}] $
         \EndFor
         \State $i^*= \arg\underset{i}{\min} \;\mathbb{E}[\Delta^{(i)}]$
         \State $P^*_{IS} = P^{(i^*)}$
    \end{algorithmic}\label{alg:IS} 
\end{algorithm}

\begin{table}[t]
    \centering
    \caption{Worst case computational complexities of the multi-source schedulers taking the maximum permitted schedule size of IS as $I$ ($I>K$).}
    \begin{tabular}{|c|c|}
    \hline
    Algorithm & Complexity \\
    \hline
    IS & $O(N^2 {I}^3)$  \\  
    \hline
    SAMS & $O(N K)$  \\   
    \hline
    Eywa & $O(N^2K^2)$ \\   
    \hline 
    \end{tabular}
    
    \label{tab:comp}
\end{table}

\subsection{Probabilistic Scheduler}
The probabilistic generate-at-will (P-GAW) model was introduced in \cite{gau_ciss24} for scheduling $N$ sources in the absence of packet drops, where at every scheduling instance,  source-$n$ is selected based on a probability $\eta_n$, such that $\sum_{n=1}^N \eta_n = 1$. In this section, we show how we can extend the results in \cite{gau_ciss24}  to find the weighted AoI of a P-GAW model in the presence of packet drops. For this purpose, we treat the system of $N$ sources with packet drops as a system with $N+1$ sources without packet drops by considering a phantom source which incorporates all the service times that resulted in a packet drop.

Let source-$(N+1)$ represent the phantom source and let $\eta'_n$ represent the probability that source-$n$ packet transmits a packet successfully (without drop). Therefore, $\eta'_n$ corresponds to the event where source-$n$ was chosen for transmission and its packet was successfully transmitted. This gives us $\eta'_n=\eta_n(1-p_n)$. Whenever a scheduled transmission fails, we assume that transmission was occupied by the phantom source. Therefore, source-$(N+1)$ transmission corresponds to events where the actual scheduled transmission has failed. This gives us $\eta'_{N+1}=\sum_{n=1}^N\eta_np_n$. The first and second moments of the service time of the phantom source are given as follows,
\begin{align}
    s_{N+1} = \frac{1}{\eta'_{N+1}}\sum_{n=1}^N\eta_np_ns_n, \qquad 
    q_{N+1} = \frac{1}{\eta'_{N+1}}\sum_{n=1}^N\eta_np_nq_n .
\end{align}
Therefore, the weighted AoI of the actual $N$-source system in the presence of packet drops, scheduled according to a P-GAW scheduler with selection probabilities $\eta_n$, is equivalent to the weighted AoI of the $(N+1)$-source system in the absence of packet drops, scheduled according to a P-GAW scheduler with selection probabilities $\eta'_n$. Hence, the AoI analysis of the P-GAW model presented in \cite{gau_ciss24} can directly be applied to this phantom-source expanded $(N+1)$-source system.

\section{Numerical Results} \label{sec:numerical}
In this section, we evaluate the performance of our constructed schedulers and benchmark them against the existing multi-source schedulers whenever possible.
We present three variations of the proposed SAMS algorithm that are implemented in our numerical experiments. In SAMS-1, the algorithm parameters are selected as $\mathcal{E}=\{ 0 \}$, $L=1$ whereas SAMS-2 employs a wider set ${\mathcal E} = \{ 0:0.2:2 \}$ for $\varepsilon$, but does not perform fixed-point iterations. Finally, SAMS-3 uses the same set ${\mathcal E}$, but employs fixed-point iterations with $L=3$. In all numerical experiments, SAMS uses the ungrouped version of the packet spreading algorithm unless specifically stated otherwise. We first consider the case when $N=2$ and compare the performance of NOTS with the optimal P-GAW scheduler denoted by P-GAW$^*$, RR (round robin) policy, IS algorithm and SAMS-3 for two sources with exponentially distributed source service times. Here, NOTS is constructed using $\alpha =50$, whereas P-GAW$^*$ is implemented by finding the optimal $\eta_n$ probabilities through exhaustive search.

\begin{figure}[t]
    \centering
    \begin{subfigure}[b]{0.49\columnwidth}
         \centering
         \includegraphics[scale=0.55]{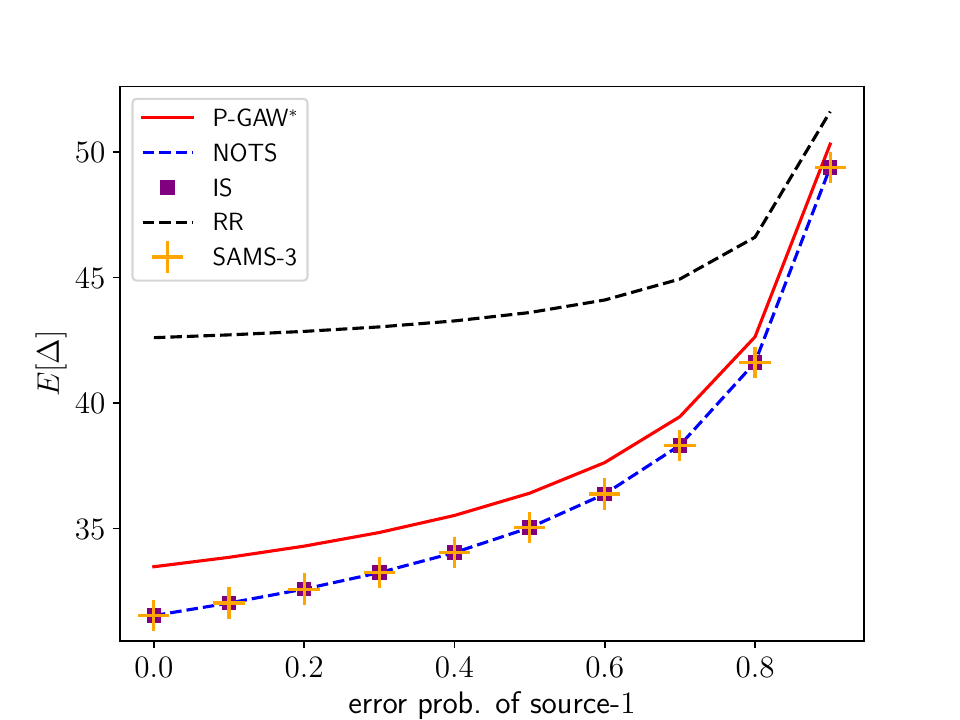}
         \caption{$s_1=2$}
     \end{subfigure} \hfill
     \begin{subfigure}[b]{0.49\columnwidth}
         \centering
         \includegraphics[scale=0.55]{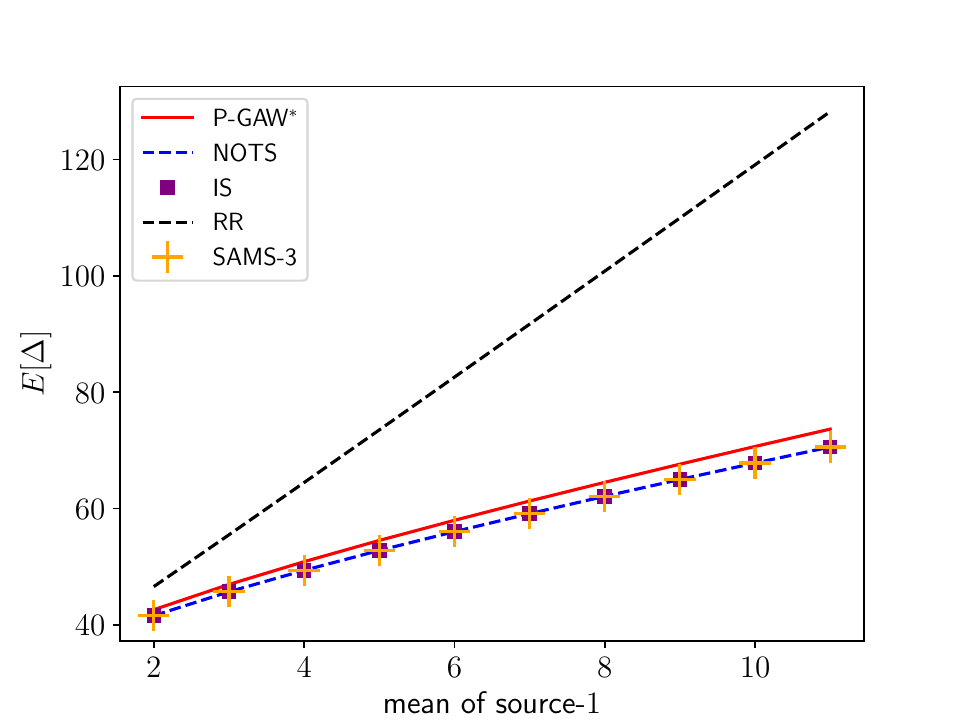}
         \caption{$p_1=0.8$}
     \end{subfigure}
    \caption{Weighted AoI of a two source system depicted as a function of (a) packet drop probability, (b) mean service time, of source-$1$, for exponential service times ($s_2=3$, $w_1=0.2$, $w_2= 0.8$, $p_2=0.9$).}
    \label{fig:exp_var_nots}
\end{figure}

\begin{figure}[th!]
    \centering
    \includegraphics[scale=0.7]{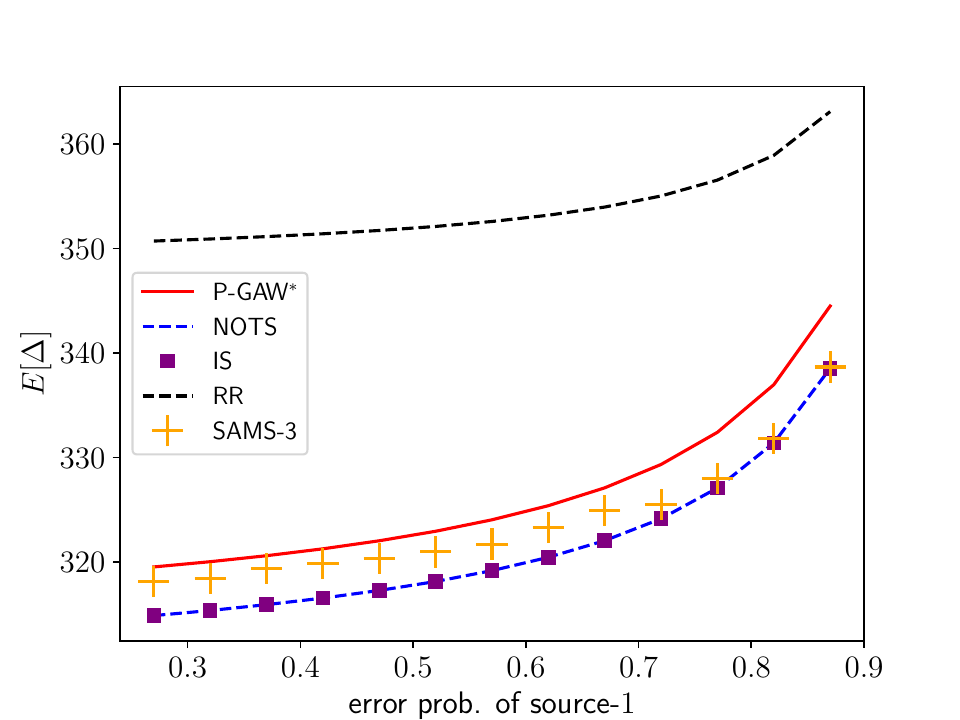}
    \caption{Variation of weighted AoI with the packet drop probability of source-$1$ for a scenario with highly variable service times ($s_1= 25$, $s_2 =24$, $c_1=2$, $c_2=15$, $p_2=0.81$, $w_1=0.04$, $w_2=0.96$).} 
    \label{fig:sams_vs_nots}
\end{figure}

In our first experiment, we fix the parameters of source-$2$ and vary either the packet drop probability or the mean service time of source-$1$. As seen in Fig.~\ref{fig:exp_var_nots},  NOTS achieves significantly lower weighted AoI than the P-GAW$^*$ and the RR policy. This is to be expected since the weighted AoI of the P-GAW model is simply an expectation of all deterministic cyclic schedules, whereas NOTS is the closest to the best cyclic schedule and is constructed by considering policies which are better than a simple RR policy.
Both IS and SAMS-3 are as good as NOTS for this particular case. However, this is not observed in all scenarios. Fig.~\ref{fig:sams_vs_nots} shows one such scenario for which there is high variability in source service times, i.e., relatively large $c_n$ values. In this case, even though SAMS-3 deviates from NOTS, IS still generates a near-optimal schedule. One of the  key reasons for the performance degradation of SAMS-3 in this scenario is the initial approximation of $\tilde{c}_n=p_n$ holding well for deterministic (or deterministic-like) service times with relatively low $c_n$ values. 

\begin{figure}[t]
    \centering
    \begin{subfigure}[b]{0.49\columnwidth}
         \centering
         \includegraphics[scale=0.55]{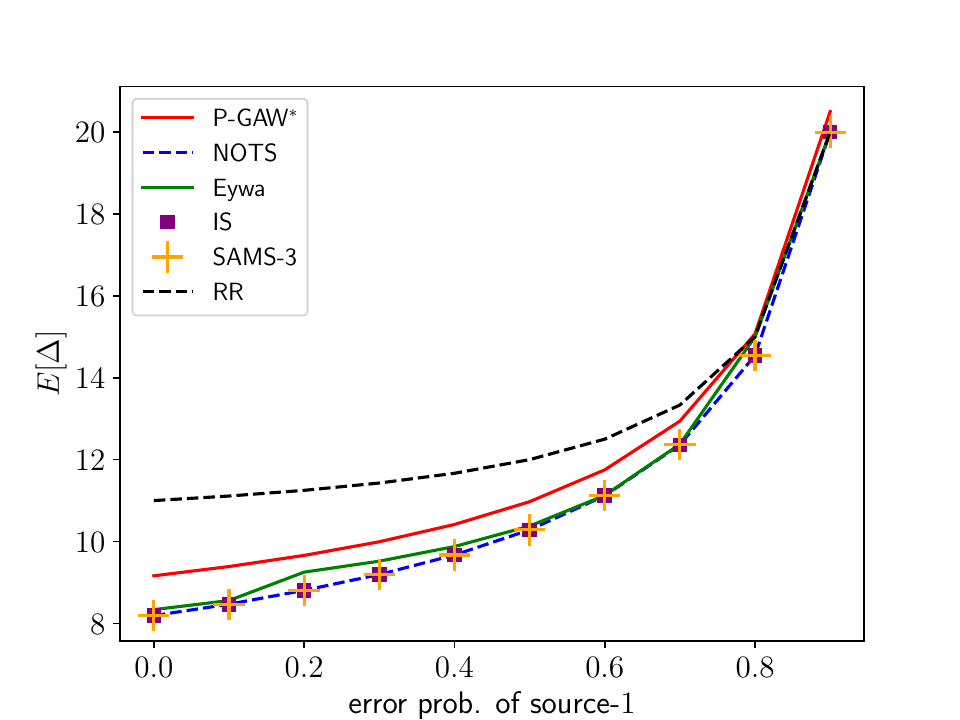}
         \caption{$s_1= s_2 =1$, $w_1=0.5$, $w_2= 0.5$, $p_2=0.9$} 
     \end{subfigure}
    \begin{subfigure}[b]{0.49\columnwidth}
         \centering
         \includegraphics[scale=0.55]{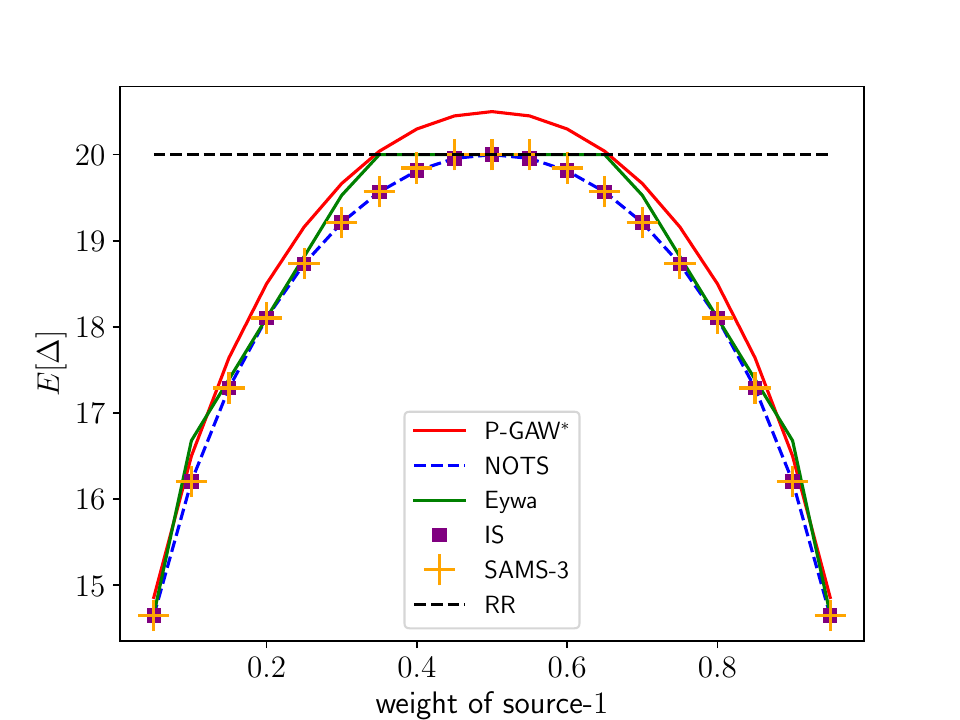}
         \caption{$s_1= s_2 =1$, $p_1=0.9$, $p_2=0.9$} 
     \end{subfigure}
    \caption{Variation of the weighted AoI of a two source system with respect to (a) packet drop probability, (b) weight, of source-$1$, for deterministic service times.}
    \label{fig:eywa_vs_nots}
\end{figure}

Next, we compare the performance of the above schedulers for identical deterministic source service times (i.e., $s_1=s_2$ and $c_1=c_2=0$). Since this experimental setup coincides with that of \emph{Eywa}, we have included \emph{Eywa} as a benchmark policy for this particular experiment. We fix the parameters of source-$2$, and in one instance, we vary the packet drop probability of source-$1$, and in the other instance, we vary the weight associated with the two sources. Fig.~\ref{fig:eywa_vs_nots} shows that NOTS outperforms \emph{Eywa} whereas SAMS-3 closely follows NOTS. \cite{gau_ciss24} shows that IS is optimal for two sources in the absence of packet errors,
and our experimental results show that IS performs very close to the optimum solution when $N=2$. However, the drawback of IS is its higher computational complexity and the lack  of provable guarantees for its performance in the presence of packet errors when $N>2$.

\begin{figure}[t]
    \centering
    \begin{subfigure}[b]{0.49\columnwidth}
         \centering
         \includegraphics[scale=0.55]{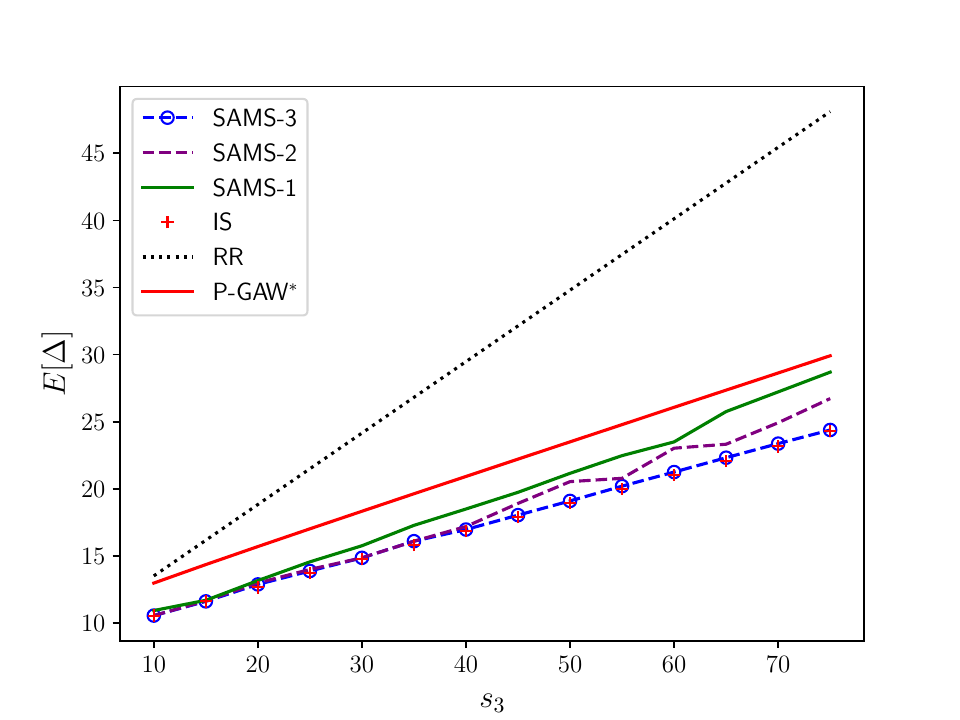}
         \caption{$s_1=5$, $s_2 =2.5$, $p_1=p_2=p_3=0$,\\ \centering$w_1=5w_2=25w_3$}  \label{fig:SAMS_N3_mean}
     \end{subfigure}
    \begin{subfigure}[b]{0.49\columnwidth}
         \centering
         \includegraphics[scale=0.55]{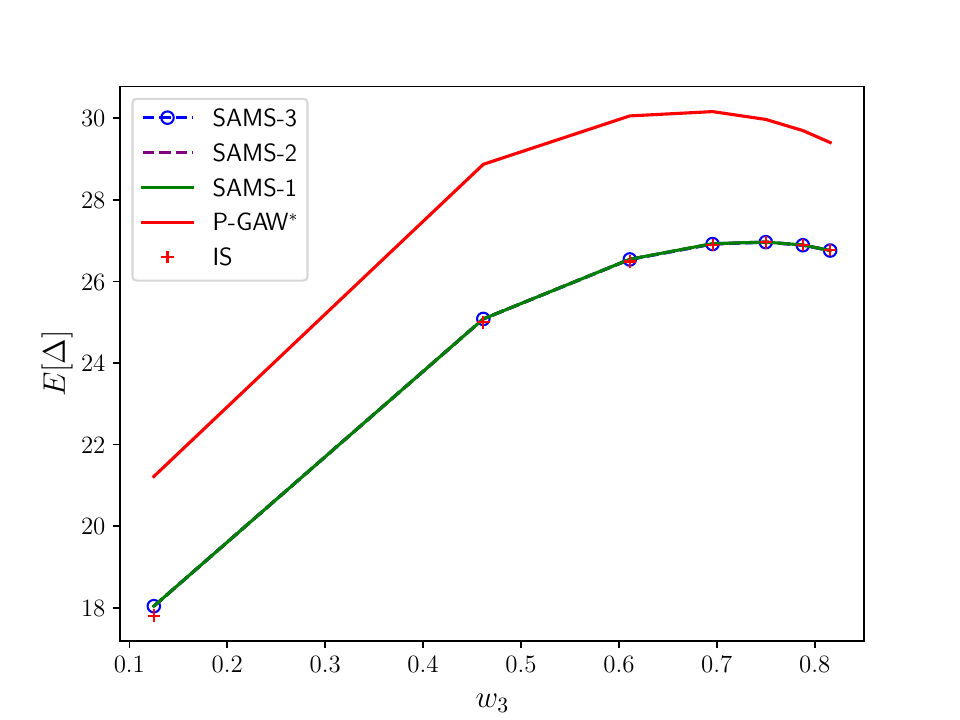}
         \caption{$s_1=10$, $s_2= s_3 =1$, $p_1=0.1$, $p_2=0.5$, \\ \centering $p_3=0.95$, $w_2=2.5w_1$} \label{fig:SAMS_N3_w}
     \end{subfigure}
    \caption{Variation of the weighted AoI of a three source system with respect to (a) packet drop probability, (b) weight, of source-$3$, for deterministic service times.}
    \label{fig:SAMS_N3}
\end{figure}

Now, we consider the case $N>2$, and compare the performance of the three variants of the  SAMS algorithm. We first study a small scale status update system with $N=3$ with deterministic service times. In this experiment, we fix the parameters of the first two sources and vary the mean and the weight of the third source. As a benchmark policy, we use the IS algorithm ($I =75$) since the above empirical evidence suggests that it is close to optimal for $N=2$. Fig.~\ref{fig:SAMS_N3_mean} depicts the variation of the weighted AoI with the mean service time  of the third source $s_3$ for a highly heterogeneous scenario with no packet drops. As illustrated, the RR policy performs very poorly especially when heterogeneity increases, i.e., $s_3$ is increased. The performance gap between P-GAW$^*$ and IS is also substantial for relatively large $s_3$ values. The results obtained with the three variations of SAMS reveal that: (i) exhaustive search over a number of $\varepsilon$ values is advantageous compared to using a single value of this parameter (observe that SAMS-2 outperforms SAMS-1); (ii) with the addition of a few fixed-point iterations, the weighted AoI performance can further be improved (observe that SAMS-3 outperforms SAMS-2). Overall, for this example, the performance of SAMS-3 has been very close to that of IS.
Fig.~\ref{fig:SAMS_N3_w} shows the variation of the weighted AoI with the weight allocated for the third source $w_3$ for a system with non-zero packet drop probabilities. The ratio of the first two sources is fixed as we vary $w_3$ in this example. In this example, all three variations of the SAMS algorithm closely follow the IS algorithm while exhibiting a significant performance gain over P-GAW$^*$. The RR policy is purposefully avoided in this example, since it is inherently worse compared to all the other policies. Note that the computational complexity of IS is $O(N^2 {I}^3)$ which is far higher than the worst case complexity of $O(N K)$ for SAMS (assuming the produced pattern size $K$ in SAMS is less than $I$), the latter being applicable to very large-scale scenarios as well. In this regard, SAMS-3 performance comes very close to IS for deterministic service times, despite the substantial gap in their computational complexities. 

\begin{figure}[t]
    \centering
    \includegraphics[scale=0.7]{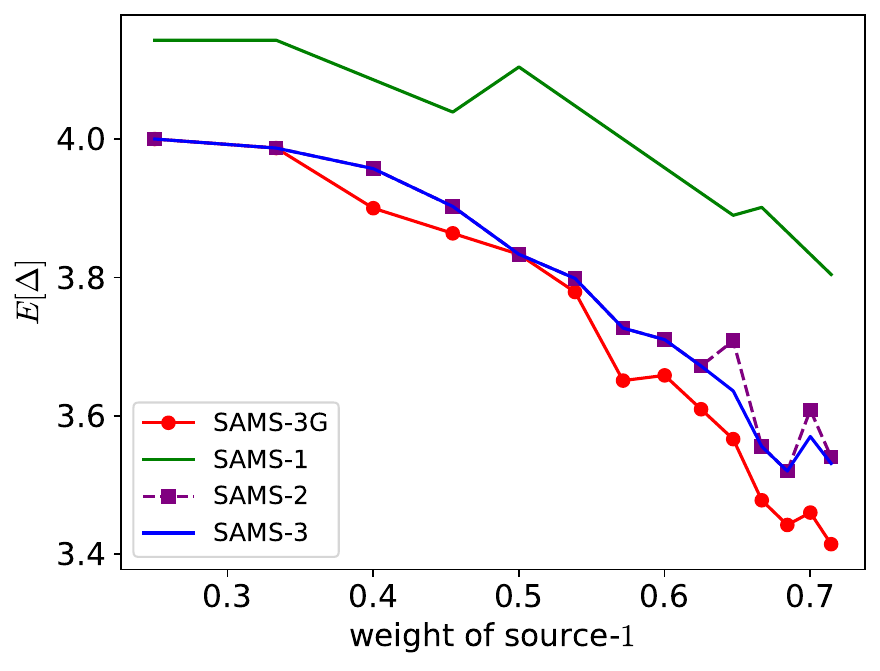}
    \caption{Variation of the weighted AoI with respect to the weight of source-$1$ for i.i.d exponentially distributed service times and zero packet error probabilities ($w_2=w_3=2w_4=2w_4$, $s_1= s_2=\dots = s_5 = 1$).}
    \label{fig:sams3G}
\end{figure}

To compare how grouping affects the packet spreading algorithm, we evaluate the performance of the SAMS-3 algorithm with its grouped counterpart, SAMS-3G. SAMS-3G uses the same parameters for $L$ and $\mathcal{E}$ that were used in SAMS-3, with the only exception being the packet spreading algorithm. As the packet spreading algorithm in SAMS-3G, we use Algorithm~\ref{alg:msDRR} as opposed to Algorithm~\ref{alg:Spreading} employed in SAMS-3. In this experiment, we consider a system with five sources with i.i.d.~exponential services times in the absence of packets errors but with different weights allocated to each source. Fig.~\ref{fig:sams3G} shows the variation of the weighted AoI with the weight of source-$1$. As shown in Fig.~\ref{fig:sams3G}, there is a clear performance gain when using the grouped packet spreading algorithm for this small-scale status update system. However, we avoid using the grouped packet spreading algorithm for large-scale status update systems due to the added complexity introduced by grouping.

\begin{figure}[t]
    \centering
    \includegraphics[scale=0.7]{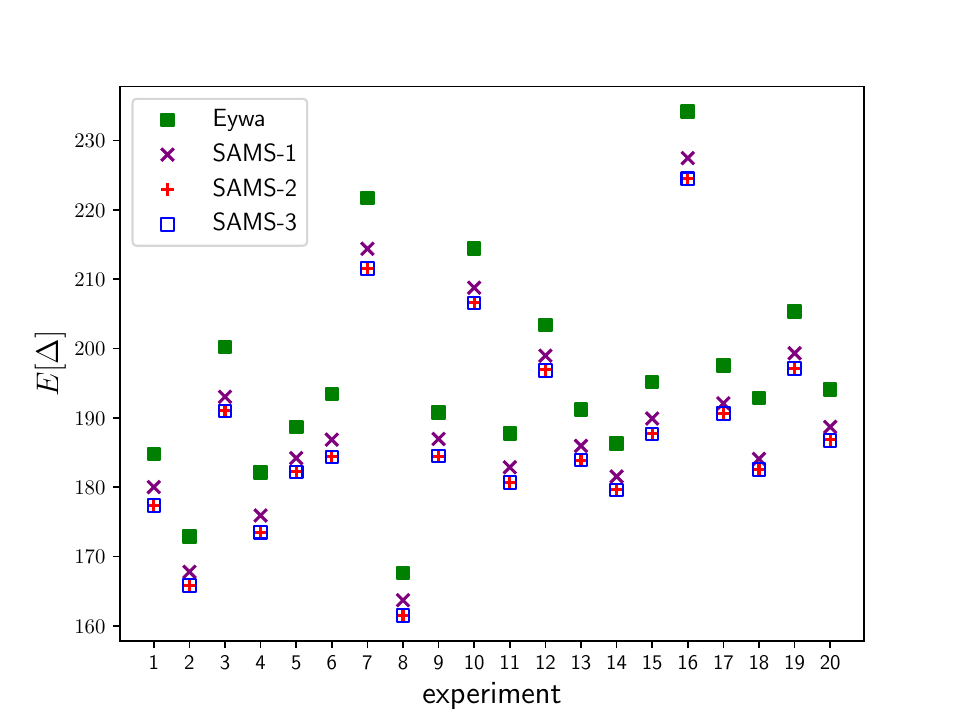}
    \caption{Weighted AoI for $N=100$ with deterministic and identical service times for twenty different experiments.}
    \label{fig:Eywa}
\end{figure}

Next, to evaluate the performance of the SAMS algorithm for large-scale status update systems. We compare the performance of SAMS against the \emph{Eywa} framework  which is designed only for systems with deterministic and identical service times. We have intentionally left out the P-GAW$^*$ scheduler for this large-scale system, since the optimal probabilities of the probabilistic scheduler is found via an exhaustive search, and therefore would be infeasible in these large-scale scenarios. To bring our model to the same domain as \emph{Eywa}, we set the service times of all sources to 1 with $c_n=0, \ \forall n$. We take $N=100$ sources, and randomly sample 20 weight vectors and probability of error vectors, and evaluate the weighted AoI of the two schemes, namely SAMS (with its three variations) and \emph{Eywa}, in each case. As illustrated in Fig.~\ref{fig:Eywa}, in each of the experiments, SAMS outperforms \emph{Eywa} by a significant margin. This performance gain is achieved without any sacrifice in the computational complexity. In fact, the computational complexity of \emph{Eywa} is $O(N^2I^2)$. Moreover, in this example, SAMS-2 and SAMS-3 performances are very close. Actually, the $\tilde{c}_n$ values obtained by the SAMS-2 scheduler are already very close to $p_n$, and therefore, there is not much need for performing fixed-point iterations specific to SAMS-3. However, SAMS-2 has a considerable performance gain over SAMS-1, indicating that even though fixed point iterations are not much useful in these large-scale systems, the local search mechanisms (finding the optimal $\varepsilon$) in SAMS-2 is advantageous. Finally, the computational complexity of \emph{Eywa} is much higher than SAMS and this is the sole reason for limiting this particular example only to 20 experiments and 100 sources.

In fact, it is quite possible to obtain well-performing schedulers for up to thousand information sources with SAMS, thanks to its low computational complexity. Now, we present the results for systems representative of massive IoT networks, i.e., $N>100$. For these experiments, we evaluate how the pattern size and the pattern computation time vary with the number of sources for massive scale (MS) scenarios named MS1, MS2, MS3 and MS4. In MS1, we consider identical deterministic service times $s_n=1$, zero packet drop probabilities $p_n=0$ and linearly increasing source weights $w_n=nw_1$. MS2 is similar to MS1 with the exception that the packet drop probabilities of the sources are evenly spaced between 0 to $0.5$ as $p_n = \frac{1}{2n}$. MS3 follows MS1 but with non-identical deterministic service times  where $s_n = n \;\text{modulo}\; 4 +1 $. Finally, MS4 uses identical exponentially distributed service times $s_n=1$, $c_n=1$, with the other parameters chosen the same as MS1. Each of these experiments is run on a Python 3 Google Compute Engine  with a 2.30GHz CPU and 12GB RAM. Table~\ref{tab:MS} summarizes the results for the four studied scenarios where we see that, even with ordinary computing resources one can obtain a schedule for $N=1024$ sources in a couple of hours. This indicates that constructing schedules for thousands of sources is well within the reach of our proposed algorithms. It is worth noting that, despite the high computational execution times observed for constructing the schedules for large number of sources, once a schedule has been constructed, 
the runtime complexity of the schedule is extremely low, i.e., $O(1)$, due to simplicity of running a cyclic scheduler.

\begin{table}[b]
    \centering
     \caption{Variation of the pattern size and the computation time in seconds of the SAMS-3 algorithm for various representative massive scale scenarios.}
    \begin{tabular}{|*{9}{c|}}
    \hline
    \multirow{2}{*}{$N$} & \multicolumn{2}{c|}{MS1}
            & \multicolumn{2}{c|}{MS2}
                    & \multicolumn{2}{c|}{MS3}
                            & \multicolumn{2}{c|}{MS4}                \\
    \cline{2-9}
    & $K$& Comp.~time  &$K$& Comp.~time  &$K$& Comp.~time  &$K$& Comp.~time \\
    \hline 
    128 & 874& 32& 1198& 71& 1041 &83 &885 &49   \\
    \hline 
    256 &1748 & 121&2384 &282 &2304 &271 & 1523 & 124  \\
    \hline 
    512 & 3283&517 & 4768& 1244&4766 & 1282&3496 & 589  \\
    \hline 
    1024 &6984 &2879 &9534 &7459 & 8297& 8833& 7047& 3475  \\
    \hline 
    \end{tabular}
   
    \label{tab:MS}
\end{table}

\section{Conclusions} \label{sec:conclusions}
In this work, we have first introduced two novel analytical frameworks to compute the weighted AoI of a multi-source status update system with $N$ sources employing a cyclic scheduler, on the basis of which we have proposed an algorithm called SAMS (scalable weighted AoI Minimization Scheduler) for constructing cyclic schedules for up to thousand sources, which can be employed in massive-size networks such as large-scale IoT networks. We have obtained a theoretical lower bound achievable by cyclic scheduling for $N=2$, and we presented an algorithm, called NOTS, that can attain this bound. The lower bound presented in this work enables us to evaluate the optimality of a number of heuristic based age-agnostic scheduling algorithms for the specific case of $N=2$ sources. For small-scale status update systems, we have shown that SAMS performs very close to the optimum solution for $N=2$, and also very close to the benchmark IS (insertion search) algorithm for deterministic service times for $N=3$. For moderately large-scale status update systems, i.e., $N=100$, SAMS is shown to perform better than the best-known age-agnostic scheduler existing in the literature that can be used for $N=100$,
with a lesser computational complexity. We have also shown that cyclic schedule construction for massive-scale status update systems with up to thousand sources is possible with the proposed SAMS algorithm, while no other existing age-agnostic scheduling algorithm appears to be feasible in this massive-connectivity regime.

Future extensions of this work may involve improved packet spreading algorithms for random service times, modeling update systems with non-colocated server and monitor, non-negligible polling delays, sources with random arrivals (RA) of packets as opposed to generate-at-will (GAW) systems, sources with duty cycle constraints, and so on. Applications of cyclic scheduling, to systems with inherent periodic characteristics such as a server taking vacations based on  a periodic schedule, is another interesting line of research. 

\bibliographystyle{unsrt}
\bibliography{refs}

\end{document}